\documentclass[journal]{IEEEtran}
\usepackage{graphicx}
\usepackage{amsmath,amssymb,amsfonts}
\usepackage{algorithm}
\usepackage{amsmath,amsthm}
\usepackage{tabularx}

\ifCLASSINFOpdf
\else
\fi
\newtheorem{remark}{Remark}
\newtheorem{theorem}{Theorem}
\newtheorem{assumption}{Assumption}
\newtheorem{pf}{Proof}

\begin{document}
%
\title{Distributed Optimal Guidance Laws for Multiple Unmanned Aerial Vehicles Attacking A Moving Target}

%
%
%

\author{Xiaoqian~Wei,

Jianying~Yang, 

        and~Xiangru~Fan,

}

%
%

\markboth{Journal of \LaTeX\ Class Files,~Vol.~14, No.~8, August~2015}%
{Shell \MakeLowercase{\textit{et al.}}: Bare Demo of IEEEtran.cls for IEEE Journals}
%



\maketitle

\begin{abstract}
In this paper, two cooperative guidance laws based on two-point boundary value are designed to deal with the problem of cooperative encirclement and simultaneous attack under condition of both known target acceleration and unknown target acceleration. The only requirement for the multi-attacker communication network is that it contains a directed spanning tree. The guidance laws can function properly as long as at least one attacker can observed the target. The acceleration components along the attacker-target line of sight in the novel guidance laws can reduce the relative remaining distance between each of the attackers and the target at the same speed, thus completing simultaneous attack and avoiding the calculation of the remaining time. The components of the guidance laws perpendicular to the attacker-target line of sight can make the normal overload of relative motion zero, so that the trajectory will be smooth and the collision problem within the attacker can be avoided. Simulation results verified the practicability of the novel guidance laws.
\end{abstract}


%
\IEEEpeerreviewmaketitle

\section{Introduction}

In recent years, the problem of multiple low-speed attackers simultaneously attacking a high-speed moving target has become a research hot spot in \cite{1,2,3,4,5}. With the research of multi-agent consistency control methods in \cite{6,7,8,9,10}, the guidance law design of multi-attackers simultaneously attacking a moving target emerges in endlessly, but it is too theoretical to consider practical problems, such as internal collision of attackers before attacking a target in \cite{11}, and multi-attackers can satisfy the optimal cost function, but the trajectory is too curved to apply to the actual situation. Even though the design of the attackers' guidance laws can make the relative motion states of the attacker-target consistent, it can not guarantee accurate attack. In addition to the method of multi-agent consistency, some researchers use the idea of attacker-target relative motion remaining time consistency to design the guidance law in
\cite{12,13}, but the remaining time can only be estimated according to the future state, and the maneuvering target can not be accurately estimated, so the current remaining time method can only deal with stationary or slowly varying average velocity targets in \cite{14}.


Based on the two-point boundary value in \cite{15} and Hamilton optimization method, this paper designs distributed cooperative guidance laws for multiple low-speed attackers to encircle and attack a high-speed moving target simultaneously. The two-point boundary value is used to determine the initial and final states of the relative motion between the attackers and the target, while Hamilton optimization makes the state of the relative motion between the attacker and the target converge to the expected value in \cite{16,17,18,19}. In these guidance laws, the relative motion of the attacker-target is decomposed into two sub-motions along the attacker-target line of sight (LOS) and perpendicular to the LOS in \cite{20,21}. In this paper, relative motion is used instead of each state to guide, so as to avoid multi-attacker collision. The sub-motion along the LOS makes the relative distances and relative velocities between the attacker and the target consistent respectively, and the remaining time of the relative motion can be achieved simultaneously and accurately without estimating. The sub-motion of vertical to LOS makes the normal overload of attacker-target relative motion converge gradually and eventually converge to zero, which will not affect the uniform motion along the LOS, but also ensure that the relative motion trajectory is smooth and has practical significance.


The main contributions of this paper are as follows: Firstly, the guidance laws in this paper are distributed, that is, attackers use the connected undirected communication network, only need the information of themselves and their neighbors, and do not need to know the information of all attackers. It should be noted that at least one attacker can observe the information of the target, while the other attackers can obtain the information of the target through the communication network and geometric relationship.


Secondly, the guidance laws in this paper can achieve accurate simultaneous hit. When the attacker arrives at the set time (terminal time), the relative distance between the attacker and the target is less than or equal to the killing radius of the attacker, and the relative distance between the attacker and the target is the same as that of other attackers. At this time, the relative velocity between the attacker and the target is the same as that of other attackers, so that the target is hit at the same time in the known time after the set time, which is the ratio of the relative distance between the attacker and the target to the relative speed at the set time. Attention should be paid to the fact that the normal overload of the attacker-target converges to a small amount, which makes the normal relative motion rotate around the target at a small speed, while the tangential relative motion becomes a uniform motion. Finally, the radius of the circle becomes smaller and smaller, and the normal overload gradually converges to zero until the precise hit.


Thirdly, the guidance laws designed in this paper are robust. When the relative motion of attacker-target is disturbed, the actual state optimization analytic solution obtained by minimizing the cost function produces errors. If the undisturbed optimization solution is continued to be used as the analytic solution, the error will accumulate. In order to deal with this situation, we use the real-time state value instead of the optimization value, and use the real-time value as the initial value to continue to calculate the analytical solution. Since the guidance laws are designed based on two-point boundary value problem and the condition of their starting and ending points is satisfied, the real-time value is used to reduce the error in the calculation process, and it can still converge to the end value.


Fourthly, the guidance laws designed in this paper are real-time. Variational method and other numerical methods used in previous research work can only design fixed trajectory on the ground first, which has a large amount of calculation and no real-time performance. The guidance laws are designed based on the two-point boundary value problem, and the analytical solution is obtained. It only changes with time, and the state at the beginning and the end of the guidance laws are determined in advance. The optimal solution path connecting the beginning and the end states is obtained by minimizing the cost function.

 The rest of this paper is organized as follows. The problem statement is given in the next section. Section \uppercase\expandafter{\romannumeral3} and Section \uppercase\expandafter{\romannumeral4} present the analyses of design of distributed guidance laws for multi-UAV cooperative attacking a moving target with known acceleration and unknown acceleration respectively based on the two-point boundary value problem. Numerical simulations are shown in Section \uppercase\expandafter{\romannumeral5}, the main contributions of the paper are summarized in Section \uppercase\expandafter{\romannumeral6}.

\vspace{-6pt}

\section{Preliminaries}
In the field of multi-agent system, the communication topology of a multi-agent system is described as a directed graph. Individual agent is treated as a vertex in the communication graph and the information communication link between two adjacent agents is modelled as an edge in the communication graph.

Take a multi-agent network consisting of N agents as an example. The directed graph $\textbf{\emph{G}}(\textbf{\emph{V}},\textbf{\emph{E}},\textbf{\emph{A}})$ is used to describe the communication topology, $\textbf{\emph{E}}\subseteq \textbf{\emph{V}}\times\textbf{\emph{V}}$ is the set of edges and the non-negative matrix $\textbf{\emph{A}}=[a_{ij}]_{N\times N}$ with elements $a_{ij}$ is the weighted adjacency matrix. An edge represents an information link between an ordered pair of nodes $(i,j)$, which stands for node j to node i, in the communication topology $\textbf{\emph{G}}$. Self-loops are not permitted in the communication topology, which means $(i,i)$ is not allowed. A directed route from node i to node j is defined as a sequence of paths, $(i,k_1)$,$(k_1,k_2)$,...,$(k_l,j)$, with different nodes $k_m, m=1,2,...,l$.

 A graph is called undirected if and only if there exists an edge $(j,i)$ in $\textbf{\emph{E}}$ for any $(i,j)\in\textbf{\emph{E}}$. This structure is equivalent to a spanning tree, which is a directed rooted tree that utilizes a directed path starting from the root vertex to connect every other vertex in the graph. When there are undirected paths between any pair of different vertices in undirected graphs, undirected graphs are considered to be connected; similarly, in directed graphs, directed graphs are considered to be strongly connected. Strongly connected graphs must contain a directed spanning tree, but not vice versa.

\section{Problem formulation}
\vspace{-2pt}
\begin{figure}[!htb]
 \centering
 \includegraphics[width=0.5\textwidth]{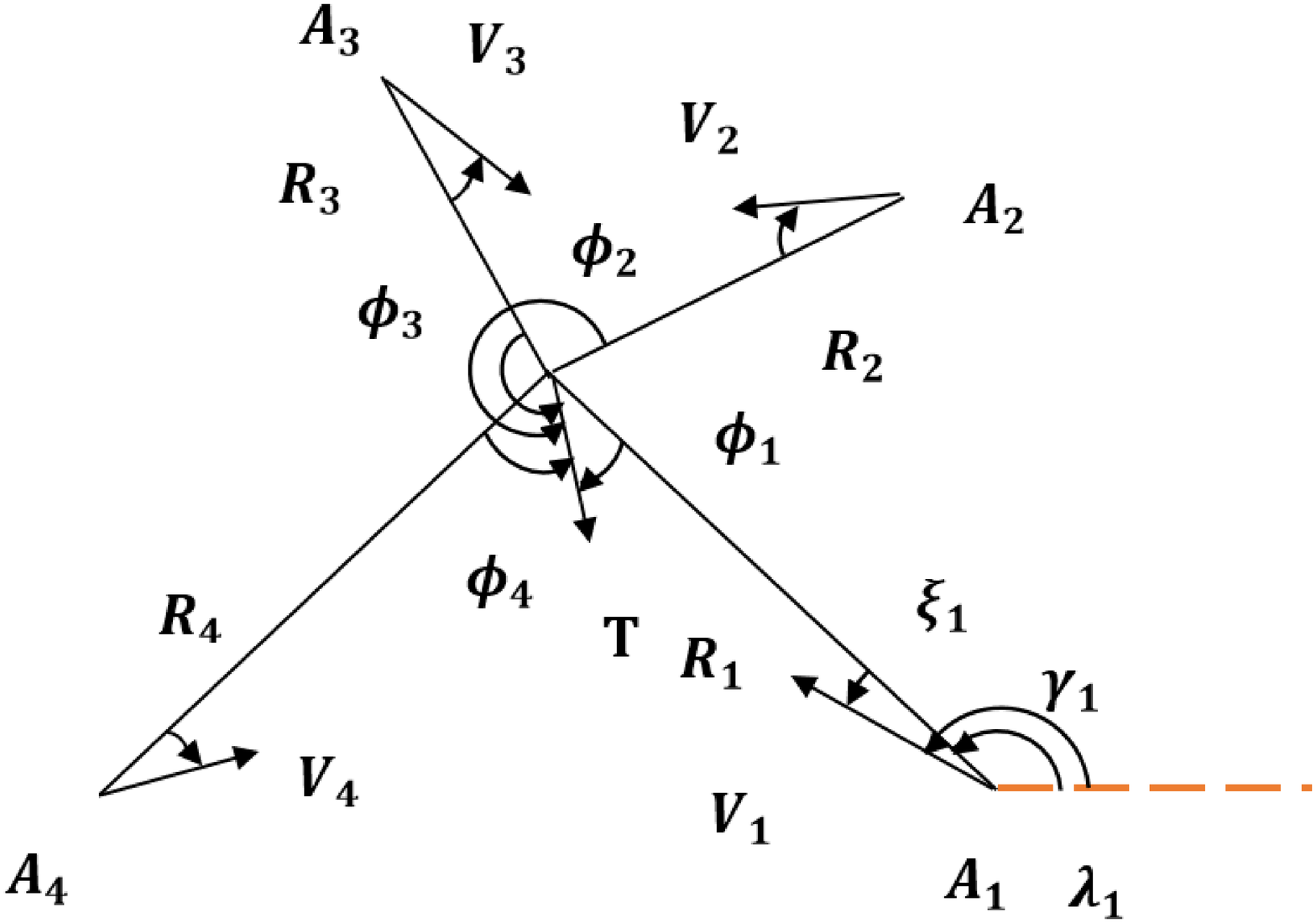}
 \caption{Geometry for TA engagement.}
 \label{Fig1}
\end{figure}

\begin{figure}[!htb]
 \centering
 \includegraphics[width=0.4\textwidth]{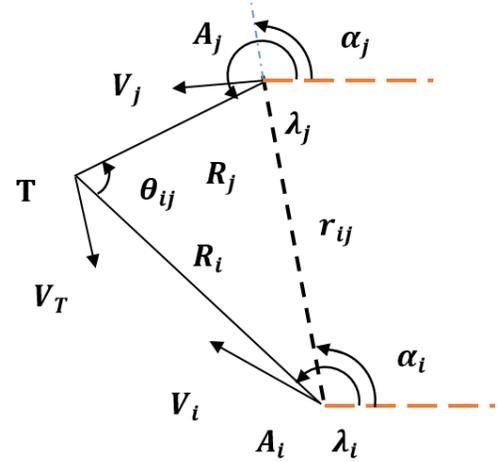}
 \caption{Geometry for an attacker and its neighbors.}
 \label{Fig2}
\end{figure}

\begin{figure}[!htb]
 \centering
 \includegraphics[width=0.45\textwidth]{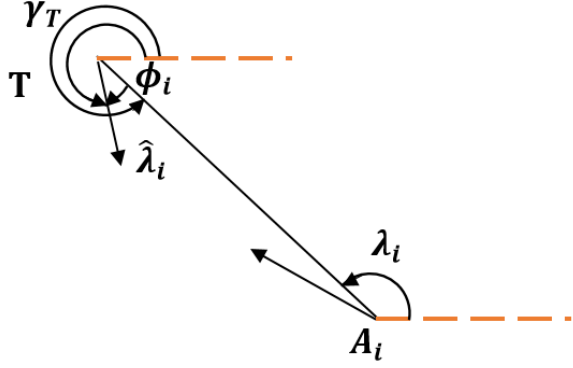}
 \caption{The relationship among angles $\delta_T$, $\phi_i$ and ${\hat{{\lambda}}_i}$.}
 \label{Fig3}
\end{figure}


In this section, we consider the TA scenario where N attackers intercept one maneuvering target in two-dimensional space in Fig.1. Multi-attackers with constant speed are denoted as nodes ${\mathcal{V}}={1,...,N}$. For simplicity, we show only a geometry for the i-th attacker and its neighbors in Fig.2. The relationship between the angles $\delta_T$, $\phi_i$ and ${\hat{{\lambda}}_i}$ is described in Figure 3. Attention should be paid to the fact that the acceleration direction of the attacker and the target is always perpendicular to their respective velocity directions, that is, their velocity values are determined ahead of time and the direction is time-varying. In this paper, the speed of the attacker is always smaller than that of the target.

The dynamic equations of the i-th attacker are given by
\begin{equation}\label{eq:1}
\begin{split}
&\dot{R_i}=V_{ri}, V_{ri}=V_T{\cos}{\phi}_{i}-V_i{\cos}{\xi}_{i}\\
&\dot{{\lambda}_i}=\frac{V_{\lambda i}}{R_i}, V_{\lambda i}=V_T{\sin}{\phi}_{i}-V_i{\sin}{\xi}_{i}\\
&{\gamma}_{i}={\xi_i}+{{\lambda}_i}, \gamma_T=\phi_i+\hat{{\lambda}}_i\\
&\dot{{\gamma}_{i}}=\frac{A_{Mi}}{V_i}, \dot{{\gamma}_{T}}=\frac{A_{T}}{V_T}\\
\end{split}
\end{equation}
with
\begin{center}
${\hat{{\lambda}}_i}$=$\left[\begin{array}{lll} {\lambda}_i-\pi & if & {\lambda}_i\geq \pi\\ {\lambda}_i+\pi & if & {\lambda}_i< \pi\end{array}\right]$
\end{center}
where the subscripts i and T denote the i-th attacker and the target. ${V_{ri}}$ are the attacker-target relative velocity components along the LOS, and ${V_{\lambda i}}$ are the attacker-target relative velocity components normal to the LOS. ${{\lambda}_i}$ is the LOS angle of the i-th attacker in the inertial reference frame, and ${\theta_{ii+1}}$ is the angle from the i-th LOS to the i+1-th LOS. The terms ${\sigma}_{i}$ and ${\sigma}_{T}$ are the heading angles of the attacker and the target. $R_i$ is the relative distance between the i-th attacker and the target, and $r_{ii+1}$ is the relative distance between the i-th and the i+1-th attacker. ${\xi}_{i}$ is the bearing angle between the i-th LOS and the direction of the i-th attacker's velocity, while the term ${\phi_i}$ is the bearing angle between the i-th LOS and the direction of the target's velocity. Positive constants $V_i$ and $V_T$ are the velocities of the i-th attacker and the target. The equations of motion for other attackers are similar.

The derivative of the above formulas are
\begin{equation}\label{eq:2}
\begin{split}
&\dot{V_{ri}}=\frac{{V^2_{\lambda i}}}{{R_i}}-{A_{Mri}}+{A_{Tri}}\\
&\dot{V_{\lambda i}}=-\frac{{V}_{\lambda i}{V}_{ri}}{R_i}-{A}_{M{\lambda}i}+{A}_{T{\lambda}i}
\end{split}
\end{equation}
where ${A_{Tri}}=-A_T\sin\phi_i$ are the target acceleration components along the LOS, while ${A_{T\lambda i}}=A_T\cos\phi_i$ are the target acceleration components normal to the LOS. ${A_{Mri}}=A_{Mi}\sin\xi_i$ are the i-th attacker's acceleration components along the LOS, while ${A_{M\lambda i}}=-A_{Mi}\cos\xi_i$ are the i-th attacker's acceleration components normal to the LOS.

\begin{assumption}\label{the1}{\rm
The graph $\textbf{\emph{G}}$ that describes the communication topology of the multi-agents system in this paper is directed and contains a spanning tree.
}
\end{assumption}

\begin{assumption}\label{the2}{\rm

}
\end{assumption}

 \begin{remark}
\label{rem1}
{\rm
On condition that the multi-attacker group fulfills Assumption 2, namely, the target velocity $V_T$, relative position $R_i$ and LOS angle $\lambda_i$ can be observed by i-th attacker, the neighbor attacker j of i-th attacker in the detection range (i.e. $j \in \boldsymbol {N}_i$) can likewise access information $V_T$, $R_i$, $\lambda_i$, the distance between itself and i-th attacker $r_{ij}$ and angle $\alpha_j$ (Notice that $\alpha_i=\alpha_j$) as illustrated in Figure 2. Then, j-th attacker can obtain the relative distance $R_j=\sqrt{R_i^2+r_{ij}^2-2R_ir_{ij}\cos(\lambda_i-\alpha_i)}$ and LOS angle $\lambda_j=\lambda_i+\arcsin(\frac{r_{ij}\sin(\lambda_i-\alpha_i)}{R_j})$ pursuant to triangle cosine theorem and triangle sine theorem. The neighbor attacker k of j-th attacker in the detection range (i.e. $k \in \boldsymbol {N}_j$) can similarly obtain the LOS angle $\lambda_k$ and relative distance $R_k$. Following Assumption 1, all attackers can get relative distance between themselves and target and LOS angles.
}
 \end{remark}

\section{Analysis}


In this segment, we discuss how low-speed attackers can coordinate enclosing or shooting high-velocity target when the acceleration of maneuvering target is known or unknown, and work out two optimal guidance laws to attack or surround simultaneously. The issue of UAV that simultaneously attacks moving a target is dealt with using target-attacker relative motion's acceleration. The target-attacker's relative motion is split into two sub-directions: tangential and normal. The normal acceleration along the LOS can make the LOS angle's angular speed converge to zero, which can prevent the early conflict that is originated from varying the LOS angle in attack's procedure. The line's tangential acceleration can manage the relative distance between multiple attackers to be the same while they enclose the target, and can likewise manage the relative velocity to achieve the equivalent negative value, namely, the relative distance declines at the equal uniform velocity, so that multiple attackers can attack the target at the same time. Pay attention that relative distances $R_i$, $R_j$, speeds of attackers $V_i$, $V_j$, and speed of target $V_T$ are positive while defined.

\subsection{Cooperative attack with known target acceleration }
In this department, we discuss how multiple attackers can encircle or attack a maneuvering target with known acceleration together.


Since the attacker's sensor can usually obtain the relative distance $R$, relative velocity $V_r$, $V_\lambda$ and relative LOS angle $\lambda$ of the target, and the position and velocity of the attacker itself are known, the attacker can calculate the position and velocity of the target at every moment, and then calculate the acceleration of the target. Although the estimation will have errors and delays which can be compensated by the observer. Therefore, we first assume that the target acceleration information is known in the theoretical study, which is consistent with most application scenarios. In future research, we will further consider how to estimate the acceleration by the relative distance, relative velocity and relative LOS angle of the target.

\begin{assumption}\label{the2}{\rm
The acceleration of the target is known.
}
\end{assumption}

\begin{theorem}\label{the1}
{\rm
For the issue of multiple attackers that encircle or attack a moving target with known acceleration at the same time in formulas~(\ref{eq:1},\ref{eq:2}) and the information transmission network satisfies Assumption 1, the acceleration of normal and tangential relative motion can be contrived as follows

\begin{equation}\label{eq:3}
\begin{split}
&{A_{Mri}}=\frac{{V^2_{\lambda i}}}{{R_i}}+{A_{Tri}}+P_{1i}^{-1}K_{1i}V_{ri}^*\\
&{A_{M\lambda i}}=-\frac{{V}_{ri}{V}_{\lambda i}}{R_i}+{{A}_{T{\lambda}i}}+P_{2i}^{-1}K_{2i}V_{\lambda i}^*\\
&V_{ri}^*=-P_1^{-1}K_1\frac{1}{N}\sum_{j=1}^Na_{ij}R_j(t_f)\exp(-P_{1i}^{-1}K_{1i}(t-t_f))\\
&V_\lambda^*=\frac{1}{N}\sum_{j=1}^Na_{ij}V_{\lambda j}(t_f)\exp(-P_{2i}^{-1}K_{2i}(t-t_f))\\
&K_{1i}=P_{1i}\frac{\ln(\frac{1}{N}\sum_{j=1}^Na_{ij}R_j(t_f))-\ln(R_i(t_0))}{t_0-t_f}\\
&K_{2i}=P_{2i}\frac{\ln(\frac{1}{N}\sum_{j=1}^Na_{ij}V_{\lambda j}(t_f))-\ln(V_{\lambda}(t_0))}{t_0-t_f}
\end{split}
\end{equation}
where $t_0$ and $t_f$ are respectively start time and end time. Note that $P_1$, $P_2$, $K_1$ and $K_2$ are positive diagonal matrices, and their diagonal elements are $P_{1i}$, $P_{2i}$, $K_{1i}$ and $K_{2i}$, $i=1,...,N$. Note that $R(t_0)=R_0$, $V_\lambda(t_0)=V_{\lambda 0}$, $R(t_f)=\frac{1}{N}\textbf{\emph{A}}R_f$ and $V_\lambda(t_f)=\frac{1}{N}\textbf{\emph{A}}V_{\lambda f}$ where $\textbf{\emph{A}}$ is the weighted adjacency matrix of attackers' information transmission network.
}
\end{theorem}

 \begin{pf}{\rm
 Our aims are to find the optimal solutions $R^*$ and $V_\lambda^*$ and their derivatives $\dot{R^*}$ and $\dot{V_\lambda^*}$ of attacker-target relative motion states $R$ and $V_\lambda$ based on the minimum cost function $J$, and to design the guidance laws $A_{Mr}$ and $A_{M\lambda}$ by using the optimal solutions and their derivatives, so that the states can be controlled by the guidance laws and the optimal solutions can be obtained at all times.

The cost function $J$ is
\begin{equation}\label{eq:4}
\begin{split}
&J=\frac{1}{2}\int_{t_0}^{t_f}(\dot{R}^TP_1^2\dot{R}+\dot{V_\lambda}^TP_2^2\dot{V_\lambda}+R^TK_1^2R+V_\lambda^TK_2^2V_\lambda)dt
\end{split}
\end{equation}

Considering that multiple attackers aim to minimize $J$, the Hamiltonian is
\begin{equation}\label{eq:5}
\begin{split}
&H=\frac{1}{2}(\dot{R}^TP_1^2\dot{R}+\dot{V_\lambda}^TP_2^2\dot{V_\lambda}+R^TK_1^2R+V_\lambda^TK_2^2V_\lambda)\\
&+\rho_{R}^T\dot{R}+\rho_{V\lambda}^T\dot{V_\lambda}
\end{split}
\end{equation}
where $\rho_{R}$ and $\rho_{V\lambda}$ are N-dimensional column vectors.

The costate dynamics are
\begin{equation}\label{eq:6}
\begin{split}
&\dot{\rho_{R}}=-\frac{\partial H}{\partial {R}}=-K_1^2R\\
&\dot{\rho_{V\lambda}}=-\frac{\partial H}{\partial {V_\lambda}}=-K_2^2V_\lambda\\
\end{split}
\end{equation}

To find the optimal inputs $\dot{R^*}$ and $\dot{V_{\lambda}^*}$ of attackers, we differentiate the Hamiltonian in $\dot{R}$ and $\dot{V_\lambda}$ and set the derivative to zero:
\begin{equation}\label{eq:7}
\begin{split}
&\frac{\partial H}{\partial \dot{R}}=P_1^2\dot{R}+\rho_{R}=0\\
&\frac{\partial H}{\partial \dot{V_{\lambda}}}=P_2^2\dot{V_\lambda}+\rho_{V\lambda}=0
\end{split}
\end{equation}

We derive quadratic derivatives of Hamilton function $H$ to prove that $\dot{R^*}$ and $\dot{V_{\lambda}^*}$ can minimize $J$.

\begin{equation}\label{eq:8}
\begin{split}
&\frac{\partial^2 H}{\partial \dot{R}^2}=P_1^2>0\\
&\frac{\partial^2 H}{\partial \dot{V_{\lambda}}^2}=P_2^2>0
\end{split}
\end{equation}


Therefore, $\dot{R^*}$ and $\dot{V_{\lambda}^*}$ can minimize $J$ to ensure that the attacker-target relative distance $R$ and the relative velocity component normal to LOS $V_\lambda$ of relative motion converge to zero, thus completing the simultaneous attack of multiple attackers on a moving target.

The boundary conditions of this control problem based on optimization are $R(t_0)=R_0$, $V_\lambda(t_0)=V_{\lambda 0}$, $R(t_f)=\frac{1}{N}\textbf{\emph{A}}R_f$ and $V_\lambda(t_f)=\frac{1}{N}\textbf{\emph{A}}V_{\lambda f}$ where $\textbf{\emph{A}}$ is the weighted adjacency matrix of attackers' information transmission network, and the final terminal condition $H(R^*(t_f),V_{\lambda}^*(t_f),\dot{R^*(t_f)},\dot{V_{\lambda}^*(t_f)},\rho_{RA}^*(t_f),\rho_{\lambda A}^*(t_f),t_f)=0$. By substituting formula~(\ref{eq:6}) into formula~(\ref{eq:7}), we obtain that

\begin{equation}\label{eq:9}
\begin{split}
&\ddot{R}=P_1^{-2}K_1^2R\\
&\ddot{V_\lambda}=P_2^{-2}K_2^2V_\lambda
\end{split}
\end{equation}

Based on boundary conditions, we design optimal solutions $R^*(t)$ and $V_\lambda^*(t)$ of relative states which satisfies formula~(\ref{eq:9}) as follows

\begin{equation}\label{eq:10}
\begin{split}
&R_i^{*}=\frac{1}{N}\sum_{j=1}^Na_{ij}R_j(t_f)\exp(-P_{1i}^{-1}K_{1i}(t-t_f))\\
&V_\lambda^{*}=\frac{1}{N}\sum_{j=1}^Na_{ij}V_{\lambda j}(t_f)\exp(-P_{2i}^{-1}K_{2i}(t-t_f))\\
&K_{1i}=P_{1i}\frac{\ln(\frac{1}{N}\sum_{j=1}^Na_{ij}R_j(t_f))-\ln(R_i(t_0))}{t_0-t_f}\\
&K_{2i}=P_{2i}\frac{\ln(\frac{1}{N}\sum_{j=1}^Na_{ij}V_{\lambda j}(t_f))-\ln(V_{\lambda i}(t_0))}{t_0-t_f}
\end{split}
\end{equation}
with $t_0<t<t_f$, $0\leq R_i(t_f)<R_i(t_0)$, and $0\leq V_{\lambda i}(t_f)<V_{\lambda i}(t_0)$.

Then, the derivatives $\dot{R^*}$ and $\dot{V_{\lambda}^*}$ of optimal solutions can be designed as follows

\begin{equation}\label{eq:11}
\begin{split}
&\dot{R^*}=-P_1^{-1}K_1R^*\\
&\dot{V_\lambda^*}=-P_2^{-1}K_2V_\lambda^*\\
\end{split}
\end{equation}

Using the acceleration components in formula~(\ref{eq:3}), we can obtain that

\begin{equation}\label{eq:12}
\begin{split}
&\dot{R}=V_r^*\\
&V_r^*=-P_1^{-1}K_1R^*\\
&\dot{V_r}=-P_1^{-1}K_1V_r^*\\
&\dot{V_\lambda}=-P_2^{-1}K_2V_\lambda^*\\
\end{split}
\end{equation}
witch means attacker-target relative states $R(t)$ and $V_\lambda(t)$ can follow the acceleration components to find their own time-varying optimal solutions and satisfy the preset boundary conditions. The proof is complete.$\qed$
 }
 \end{pf}

  \begin{remark}
\label{rem2}
{\rm

Supposing that only the acceleration component $A_{M\lambda}$ that is vertical to the LOS is managed, that is, the LOS angular speed $\dot{\lambda}=\frac{V_\lambda}{R}$ approximates zero, it can not make multiple attackers encircle or cooperatively attack the target. At this time, the LOS angular speed approximates zero (i.e., the normal overload $\dot{V_\lambda}$ approximates zero), which signifies that each attacker's LOS angle will not vary significantly before finishing the encirclement or attack task, and the trajectory is smooth, therefore effectively preventing multiple attackers' internal collision trouble ahead. It is significant to mention that multiple attackers do not hit the target necessarily at the same time, or the enclosure region is not ring-shaped.


Provided that just the acceleration component $A_{Mr}$ along the LOS is handled, in other words, the remaining distance between the attacker and the target along the LOS $R(t_f)\leq R_c$ with the killing radius $R_c$ is consistent, and all attackers' remaining distances go down at the equivalent velocity $V_r(t_f)$, so that multiple attackers can encircle (the encircling region is ring-shaped) or together attack the target at $T=t_f+t_c, t_c=|\frac{R(t_f)}{V_r(t_f)}|$. It ought to be mentioned that the normal overload $\dot{V_\lambda}(t_f)$ is not of necessity zero, multiple attackers probably collide beforehand, and the trajectory is not necessarily smooth and may not own realistic significance.

In conclusion, both the normal and tangential acceleration components of the attacker-target relative motion along LOS should be controlled, the task of multiple attackers encircling or attacking a moving target simultaneously will be finished.
}
 \end{remark}

  \begin{remark}
\label{rem3}
{\rm
The guidance laws obtain the convergence time from the initial and final values of the states, and can complete the multi-attacker simultaneous attack task with fixed or limited time.


The guidance laws are designed based on the minimization cost function $J$, which means that the relative motion states of the attacker-target will obtain the optimal values at all times according to the guidance laws. The coefficient matrices $P_1^{-1}K_1$ and $P_2^{-1}K_2$ of the exponential functions are determined by the information of the beginning and the end of the states. When $t_0<t<t_f$, the optimal solutions of the states are time-varying and always converge to the terminal values of the states. The guidance laws contain the optimal values of states, which makes the guidance laws approximate to closed-loop feedback control inputs and have robustness. The terminal value of the state $R$ converges to less than the killing radius $R_c>0$ and achieves consistency. At this time, the relative velocity components $V_r$ along the LOS is also consistent, and the normal overload $\dot{V_\lambda}$ converges to zero, then the relative motion along the LOS is uniform, and attackers can hit the target at the same time.
}
 \end{remark}

  \begin{remark}
\label{rem4}
{\rm
The guidance laws in this article are the attacker-target relative movement's acceleration components. Their essence are improved proportional guidance laws, namely, the first item of ${A_{M\lambda i}}$ in formula~(\ref{eq:3}) is proportionate to $\dot{\lambda_i}$ and the proportion is $-{V}_{ri}$, and the other items can be regarded as correction items.
}
 \end{remark}

 \subsection{Cooperative attack with unknown target acceleration}
In this department, we discuss how multiple attackers can encircle or attack a maneuvering target with unknown acceleration simultaneously. In this section, the target acceleration is unknown, but the change structure of the target acceleration is known, and its initial condition is unknown. Therefore, the current acceleration of the target is unknown. This is a common method to deal with an unknown target, which has certain practical significance. Although the maneuvering information of the actual target is unknown, the type of the target (such as aircraft or motor vehicles) is known. Therefore, we can assume that the basic characteristic structure of its motion is known. This approach is also in line with most of the actual situation.


Assuming that

\begin{equation}\label{eq:13}
\begin{split}
&\dot{A_T}=sA_T
\end{split}
\end{equation}
which means it is an exogenous system, where $s\leq0$ is a known constant. It should be noted that when $s$ is greater than zero, that is, the acceleration of the target does not converge to an upper bound with the increase of time. This situation seldom exists in practice because it does not meet the physical limits.

Distributed disturbance observer is designed in the following form
\begin{equation}\label{eq:14}
\begin{split}
&\dot{z_i}=sz_i+\sigma_{1i}V_{\lambda i}+\sigma_{2i}V_{ri}\\
&z_i=\hat{A_{Ti}}
\end{split}
\end{equation}
where $\hat{A_{Ti}}$ is the target acceleration observed by i-th attacker. $\sigma_{1i}$ and $\sigma_{2i}, i \in \textbf{\emph{V}}={1,2,...,N}$ are the observer coefficients to be determined later, and $z_i$ is the virtual state of the disturbance observer. When the guidance law is designed with the optimal values $R^*$, $V_r^*=\dot{R^*}$, $V_\lambda^*$ and $\dot{V_\lambda^*}$ satisfying the minimization of the optimization function $J$, the design of the observer likewise uses the optimal values. The following shows that selecting the right $\sigma_{1i}$ and $\sigma_{2i}, i \in \textbf{\emph{V}}={1,2,...,N}$ allows the realization of consistency of $R_i$, $V_{ri}$ and $V_{\lambda i}$, $\forall i, j \in \textbf{\emph{V}}={1,2,...,N}$.

\begin{theorem}\label{the2}
{\rm
For the problem of multiple attackers encircling or attacking a moving target whose acceleration is in the form of equation~(\ref{eq:13}) simultaneously in formulas~(\ref{eq:1},\ref{eq:2}) and the information transmission network satisfies Assumption 1, the acceleration of normal and tangential relative motion can be designed in equation~(\ref{eq:15}) along with the distributed disturbance observer in equation~(\ref{eq:14}) whose observer coefficients satisfy $\sigma_{1i}=\cos{\phi}_{i}$ and $\sigma_{2i}=-\sin{\phi}_{i}, i \in \textbf{\emph{V}}={1,2,...,N}$.

\begin{equation}\label{eq:15}
\begin{split}
&{A_{M\lambda i}}=-\frac{{V}_{ri}{V}_{\lambda i}}{R_i}+\hat{A_{T\lambda i}}+P_{1i}^{-1}K_{1i}V_{ri}^*\\
&{A_{Mri}}=\frac{{V^2_{\lambda i}}}{{R_i}}+\hat{A_{Tri}}+P_{2i}^{-1}K_{2i}V_{\lambda i}^*\\
&V_{ri}^*=-P_1^{-1}K_1\frac{1}{N}\sum_{j=1}^Na_{ij}R_j(t_f)\exp(-P_{1i}^{-1}K_{1i}(t-t_f))\\
&V_\lambda^*=\frac{1}{N}\sum_{j=1}^Na_{ij}V_{\lambda j}(t_f)\exp(-P_{2i}^{-1}K_{2i}(t-t_f))\\
&K_{1i}=P_{1i}\frac{\ln(\frac{1}{N}\sum_{j=1}^Na_{ij}R_j(t_f))-\ln(R_i(t_0))}{t_0-t_f}\\
&K_{2i}=P_{2i}\frac{\ln(\frac{1}{N}\sum_{j=1}^Na_{ij}V_{\lambda j}(t_f))-\ln(V_{\lambda}(t_0))}{t_0-t_f}
\end{split}
\end{equation}
where $\hat{A_{T\lambda i}}=\hat{A_{Ti}}\cos\phi_i$ and $\hat{A_{Tri}}=-\hat{A_{Ti}}\sin\phi_i$. $t_0$ and $t_f$ are respectively start time and end time. Note that $P_1$, $P_2$, $K_1$ and $K_2$ are positive diagonal matrices, and their diagonal elements are $P_{1i}$, $P_{2i}$, $K_{1i}$ and $K_{2i}$, $i=1,...,N$. Note that $R(t_0)=R_0$, $V_\lambda(t_0)=V_{\lambda 0}$, $R(t_f)=\frac{1}{N}\textbf{\emph{A}}R_f$ and $V_\lambda(t_f)=\frac{1}{N}\textbf{\emph{A}}V_{\lambda f}$ where $\textbf{\emph{A}}$ is the weighted adjacency matrix of attackers' information transmission network.
}
\end{theorem}

 \begin{pf}{\rm
First, we utilize the cost function $J$ and the Hamiltonian in formulas~(\ref{eq:4},\ref{eq:5}) to obtain the optimal solutions that minimize $J$. Similar to the formulas~(\ref{eq:6},\ref{eq:7},\ref{eq:8}), the optimal solutions of ${R^*}$ and ${V_{\lambda}^*}$ which satisfy the preset boundary conditions and their derivatives $\dot{R^*}$ and $\dot{V_{\lambda}^*}$ can be obtained as follows.

\begin{equation}\label{eq:16}
\begin{split}
&R_i^*=\frac{1}{N}\sum_{j=1}^Na_{ij}R_j(t_f)\exp(-P_{1i}^{-1}K_{1i}(t-t_f))\\
&V_\lambda^*=\frac{1}{N}\sum_{j=1}^Na_{ij}V_{\lambda j}(t_f)\exp(-P_{2i}^{-1}K_{2i}(t-t_f))\\
&\dot{R^*_i}=-P_{1i}^{-1}K_{1i}R^*_i(t)\\
&\dot{V_{\lambda i}^*}=-P_{2i}^{-1}K_{2i}V_{\lambda i}^*(t)\\
&K_{1i}=P_{1i}\frac{\ln(\frac{1}{N}\sum_{j=1}^Na_{ij}R_j(t_f))-\ln(R_i(t_0))}{t_0-t_f}\\
&K_{2i}=P_{2i}\frac{\ln(\frac{1}{N}\sum_{j=1}^Na_{ij}V_{\lambda j}(t_f))-\ln(V_{\lambda i}(t_0))}{t_0-t_f}
\end{split}
\end{equation}
with $t_0<t<t_f$, $0\leq R_i(t_f)<R_i(t_0)$, and $0\leq V_{\lambda i}(t_f)<V_{\lambda i}(t_0)$.

When the target's acceleration is unknown, the guidance laws based on acceleration observation error $\tilde{A_T}=A_T-\hat{A_T}$ and optimal values ${R^*}$, ${V_{\lambda}^*}$, $\dot{R^*}$ and $\dot{V_{\lambda}^*}$ are shown in formula~(\ref{eq:15}), and the derivatives of corresponding states are

\begin{equation}\label{eq:17}
\begin{split}
&\dot{V^*_{ri}}=\tilde{A_{Tri}}-P_{1i}^{-1}K_{1i}V^*_{ri}\\
&\dot{V_{\lambda i}^*}=\tilde{A_{T\lambda i}}-P_{2i}^{-1}K_{2i}V_{\lambda i}^*
\end{split}
\end{equation}
with $\tilde{A_{Tri}}=A_{Tri}-\hat{A_{Tri}}$ and $\tilde{A_{T\lambda i}}=A_{T\lambda i}-\hat{A_{T\lambda i}}$.

Then, we use a Lyapunov function $V$ to prove that the observation errors of acceleration $\hat{A_{Tr}}$ and $\hat{A_{T\lambda}}$ will not affect the convergence of the states.

\begin{equation}\label{eq:18}
\begin{split}
&V=\frac{1}{2}(R^{*T}R^*+V_r^{*T}V_r^*+V_\lambda^{*T}V_\lambda^*+\tilde{A_{T}}^T\tilde{A_{T}})
\end{split}
\end{equation}

The derivative of time for function $V$ can be obtained as follows.

\begin{equation}\label{eq:19}
\begin{split}
&\dot{V}=R^{*T}\dot{R^*}+V_r^{*T}\dot{V_r^*}+V_\lambda^{*T}\dot{V_\lambda^*}+\tilde{A_{T}}^T\dot{\tilde{A_{T}}}\\
&=R^{*T}\dot{R^*}+V_r^{*T}(\tilde{A_{Tr}}+P_{1}^{-2}K_{1}^2R^*)+V_\lambda^{*T}(\tilde{A_{T\lambda}}-P_{2}^{-1}K_{2}V_{\lambda}^*)\\
&+\tilde{A_{T}}^T(s\tilde{A_{T}}-\sigma_{1}V_r^*-\sigma_{2}V_\lambda^*)\\
&=V_r^{*T}(-\sin\phi-\sigma_{1})\tilde{A_T}+V_\lambda^{*T}(\cos\phi-\sigma_{2})\tilde{A_T}\\
&-V_\lambda^{*T}P_2^{-1}K_2V_{\lambda}^*+s\tilde{A_T}^T\tilde{A_T}+V_r^{*T}(I_N+P_1^{-2}K_1^2)R^*
\end{split}
\end{equation}
where $\sigma_{1}=diag(\sigma_{1i})$, $\sigma_{2}=diag(\sigma_{2i})$ and $\phi=diag(\phi_i)$ are diagonal matrices. $\boldsymbol{I_N}$ is a N-dimensional unit matrix.

Note that observer coefficients satisfy $\sigma_{1i}=\cos{\phi}_{i}$ and $\sigma_{2i}=-\sin{\phi}_{i}$, and conditions $V_r^*=\dot{R^*}=-P_1^{-1}K_1R^*$ and $s\leq0$, we can obtain that

\begin{equation}\label{eq:20}
\begin{split}
&\dot{V}\leq-V_\lambda^{*T}P_2^{-1}K_2V_{\lambda}^*-R^{*T}P_1^{-1}K_1(I_N+P_1^{-2}K_1^2)R^*<0
\end{split}
\end{equation}
with positive diagonal matrices $P_1$, $P_2$, $K_1$ and $K_2$.

Therefore, using the guidance law in formula~(\ref{eq:15}), multiple attackers can attack a moving target whose acceleration is unknown at the same time.$\qed$

 }
 \end{pf}

  \begin{remark}
\label{rem5}
{\rm

In formula~(\ref{eq:15}), the design of guidance law only uses the initial and final values of the relative state between multiple attackers and a target. The guidance law is open-loop, which can not eliminate external interference, resulting in poor robustness. If the following piecewise guidance law of $t\in[t_k, t_{k+1}]$ is adopted, the attack process can be controlled piecewise, so that information exchange between multiple attackers can be carried out, and the relative state of multiple attackers and the target at the time $t_{k+1}$ can be averaged of relative state at $t_k, k=0,1,\cdots,f-1,f$. When the relative velocity of the vertical LOS is zero, the relative distance between the multi-attacker and the target will be reduced to zero at time $T=\sum_{k=0}^{f-1}(t_{k+1}-t_k+|\frac{R(t_k)}{V_r(t_k)}|)$, thus completing the simultaneous attack of multiple attackers against the target. The guidance law has a closed-loop form, which enhances robustness. 

\begin{equation}\label{eq:21}
\begin{aligned}
&{A_{M\lambda i}}=-\frac{{V}_{ri}{V}_{\lambda i}}{R_i}+\hat{A}_{T\lambda i}+P_{1i}^{-1}K_{1i}V_{ri}^*,\\
&{A_{Mri}}=\frac{{V^2_{\lambda i}}}{{R_i}}+\hat{A}_{Tri}+P_{2i}^{-1}K_{2i}V_{\lambda i}^*,\\
&V_{ri}^*=-P_{1i}^{-1}K_{1i}\frac{1}{N}\sum_{j=1}^Na_{ij}R_j(t_k)\exp(-P_{1i}^{-1}K_{1i}(t-t_k)),\\
&V_{\lambda i}^*=\frac{1}{N}\sum_{j=1}^Na_{ij}V_{\lambda j}(t_k)\exp(-P_{2i}^{-1}K_{2i}(t-t_k)),\\
&K_{1i}=P_{1i}\frac{\ln(\frac{1}{N}\sum_{j=1}^Na_{ij}R_j(t_k))-\ln(R_i(t_{k}))}{t_k-t_{k+1}},\\
&K_{2i}=P_{2i}\frac{\ln(\frac{1}{N}\sum_{j=1}^Na_{ij}V_{\lambda j}(t_k))-\ln(V_{\lambda i}(t_{k}))}{t_k-t_{k+1}},
\end{aligned}
\end{equation}
where $\hat{A}_{T\lambda i}=\hat{A}_{Ti}\cos\phi_i$ and $\hat{A}_{Tri}=-\hat{A}_{Ti}\sin\phi_i$. $t_k$ and $t_{k+1}$ are the initial and end times of $t\in[t_k, t_{k+1}]$, respectively.  $P_1$, $P_2$, $K_1$ and $K_2$ are positive definite diagonal matrices with diagonal elements of $P_{1i}$, $P_{2i}$, $K_{1i}$ and $K_{2i}$, $i=1,\cdots,N$, respectively. The starting and ending boundary conditions of the relative state are $R_k=R(t_k)$, $V_{\lambda k}=V_\lambda(t_k)$, $R(t_{k+1})=\frac{1}{N}\bar{A}R_k$ and $V_\lambda(t_{k+1})=\frac{1}{N}\bar{A}V_{\lambda k}$, where $\bar{A}$ is the adjacency matrix of the attacker's information transfer topology. The proof of this sufficient condition is similar to that of guidance law~(\ref{eq:15}).


}
 \end{remark}

 \section{Simulation Results}
 In this segment, to prove the proposed guidance laws' effectiveness, numeric simulations for a multi-UAV simultaneous attack with known or unknown target acceleration are organized. Initial parameters are listed in Table 1 and Table 2.

\subsection{Example 1: Cooperative attack with known target acceleration}

\begin{figure}[!hbt]
\centering
\includegraphics[width=3in]{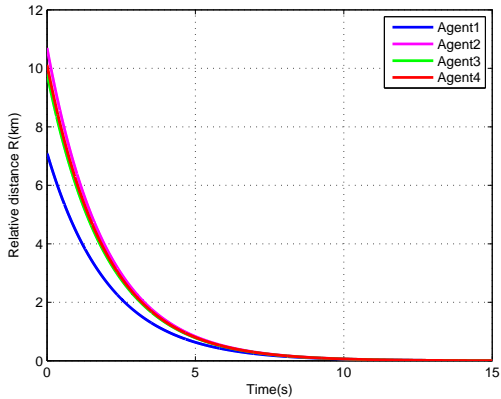}
\caption{Relative distances.\label{Fig4}}
  \label{Fig4}
\end{figure}

\begin{figure}[!hbt]
\centering
\includegraphics[width=3in]{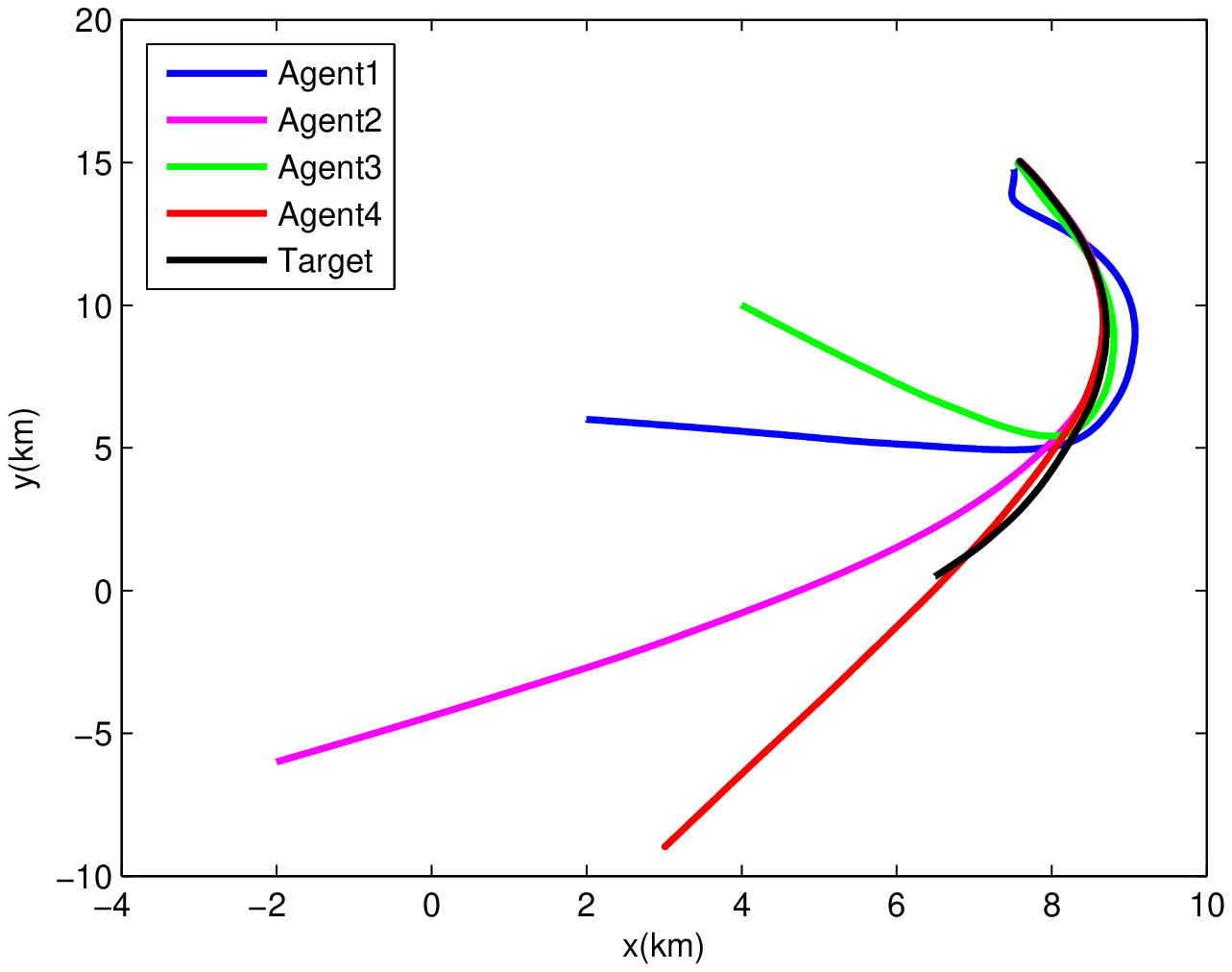}
\caption{Trajectories.\label{Fig5}}
  \label{Fig5}
\end{figure}

\begin{figure}[!hbt]
\centering
\includegraphics[width=3in]{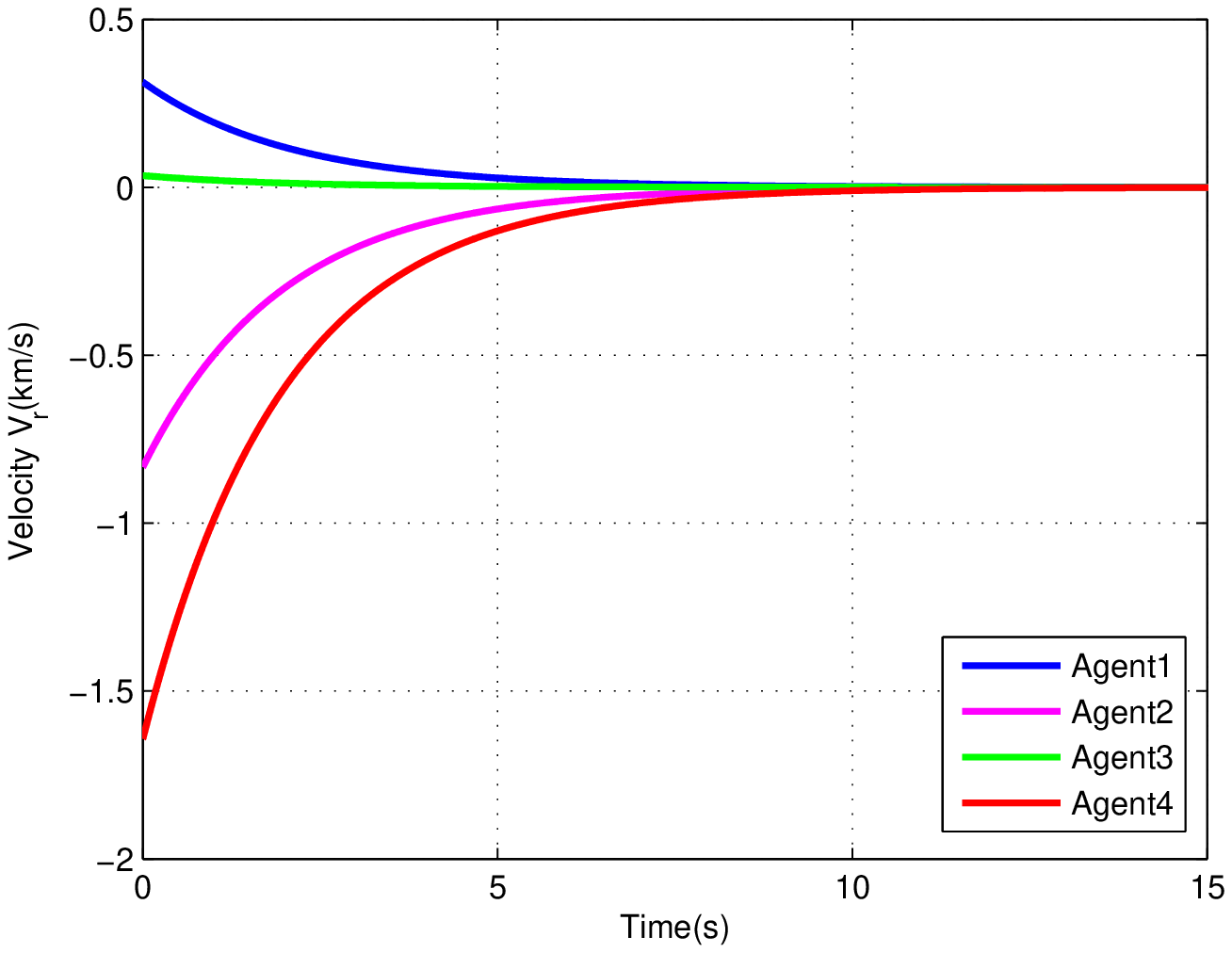}
\caption{Velocities ${V_r}$.\label{Fig6}}
  \label{Fig6}
\end{figure}

\begin{figure}[!hbt]
\centering
\includegraphics[width=3in]{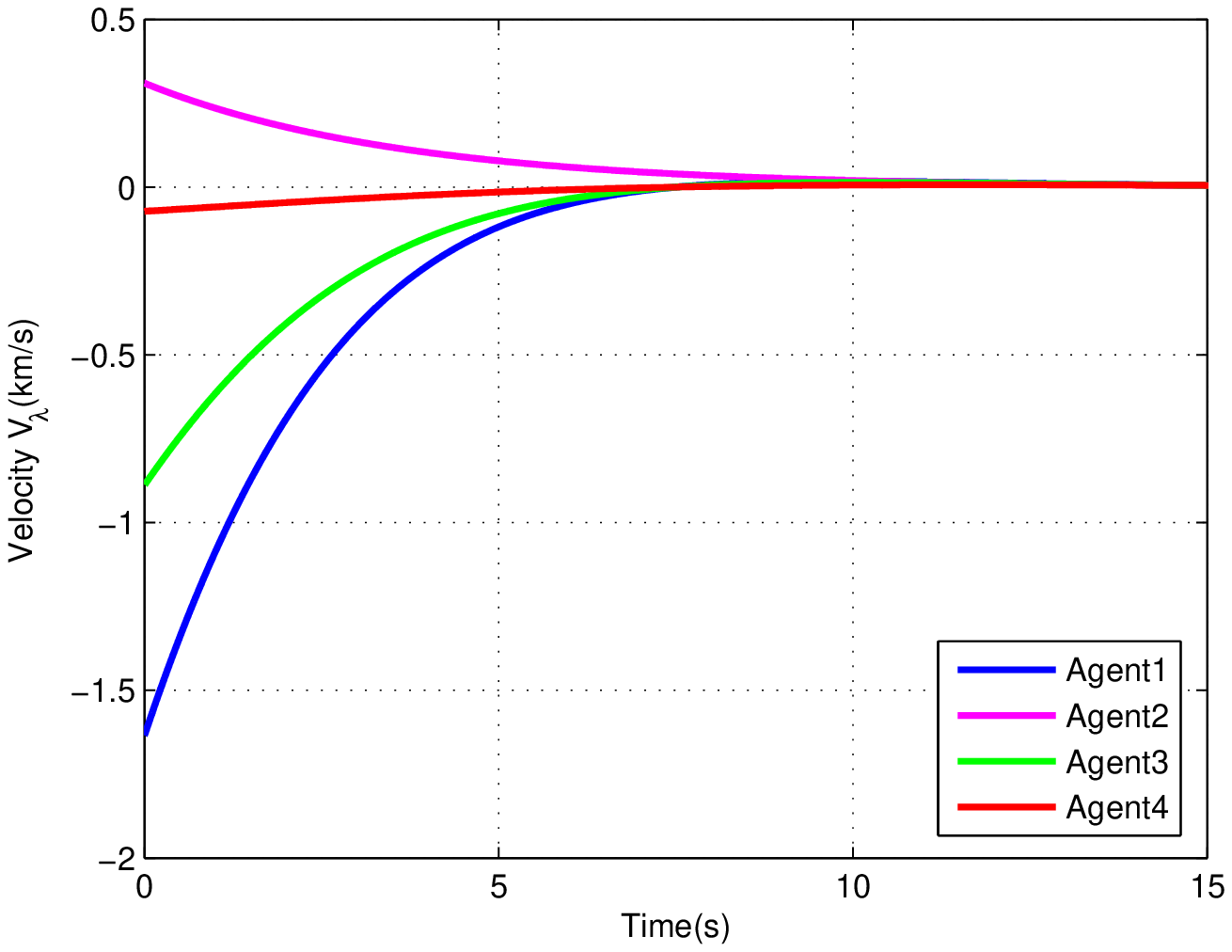}
\caption{Velocities ${V_\lambda}$.\label{Fig7}}
  \label{Fig7}
\end{figure}

\begin{figure}[!hbt]
\centering
\includegraphics[width=3in]{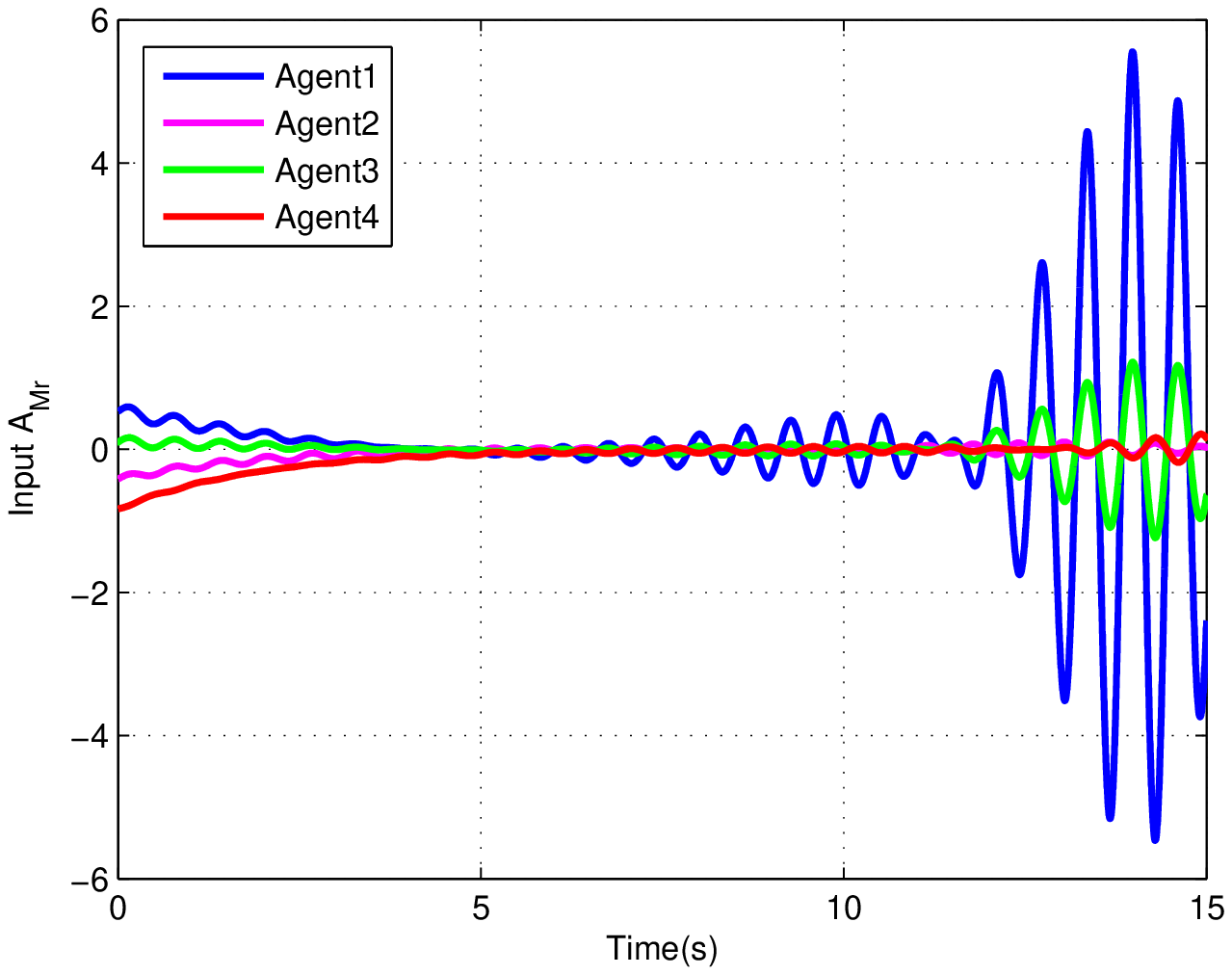}
\caption{Inputs ${A_{Mr}}$.\label{Fig8}}
  \label{Fig8}
\end{figure}

\begin{figure}[!hbt]
\centering
\includegraphics[width=3in]{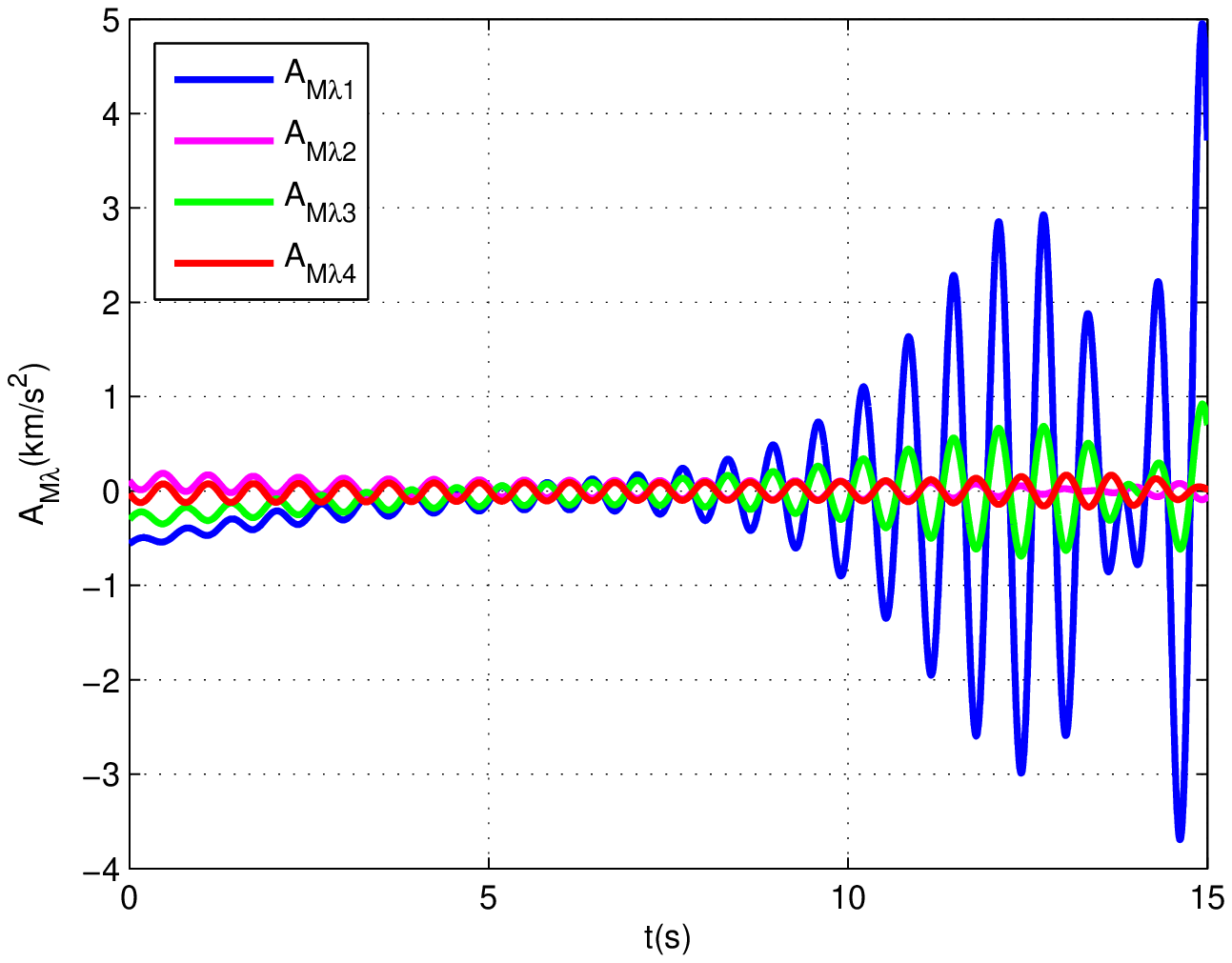}
\caption{Inputs ${A_{M\lambda}}$.\label{Fig9}}
  \label{Fig9}
\end{figure}

\begin{figure}[!hbt]
\centering
\includegraphics[width=3in]{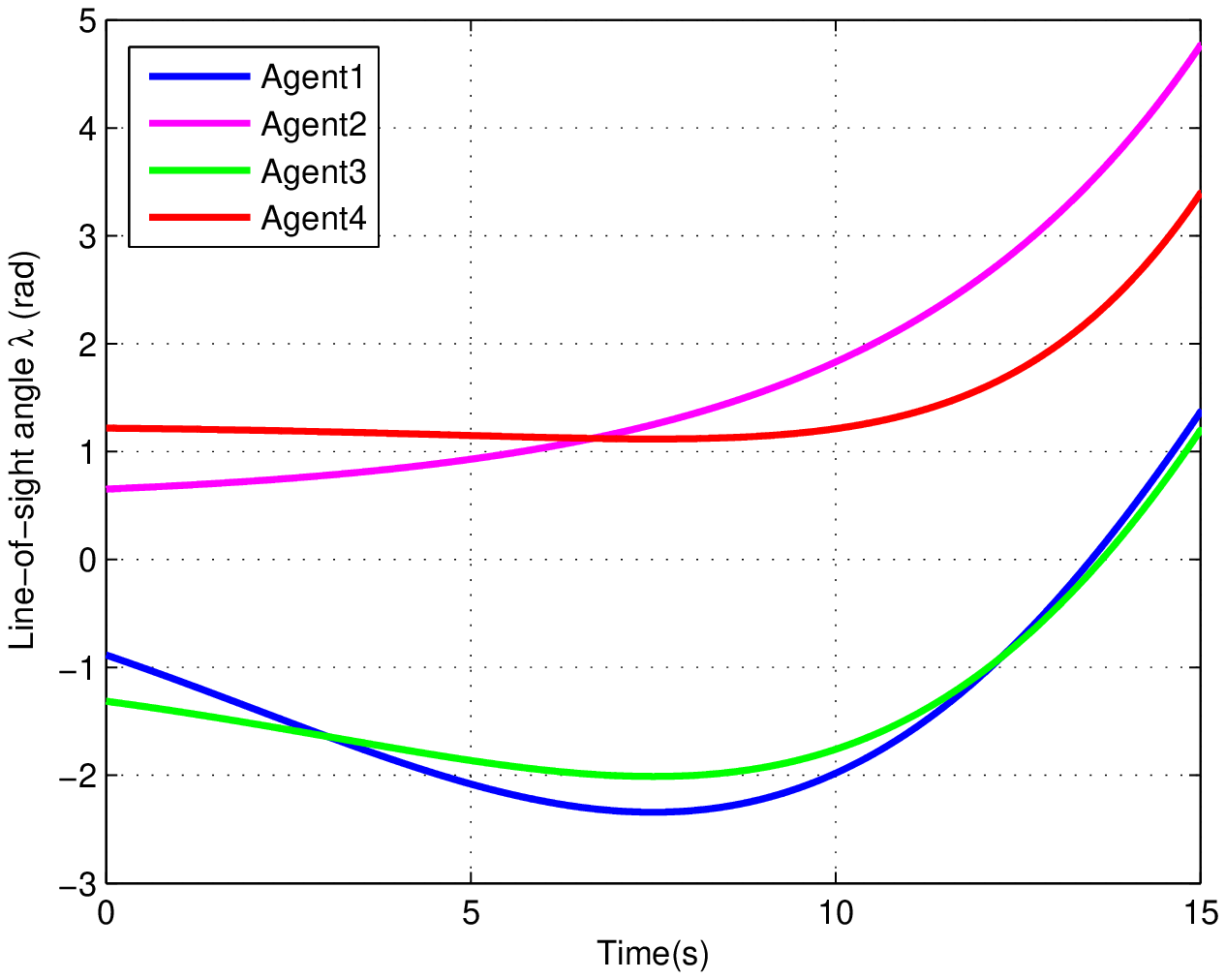}
\caption{Line of sight angle ${\lambda}$.\label{Fig10}}
  \label{Fig10}
\end{figure}

\begin{figure}[!hbt]
\centering
\includegraphics[width=3in]{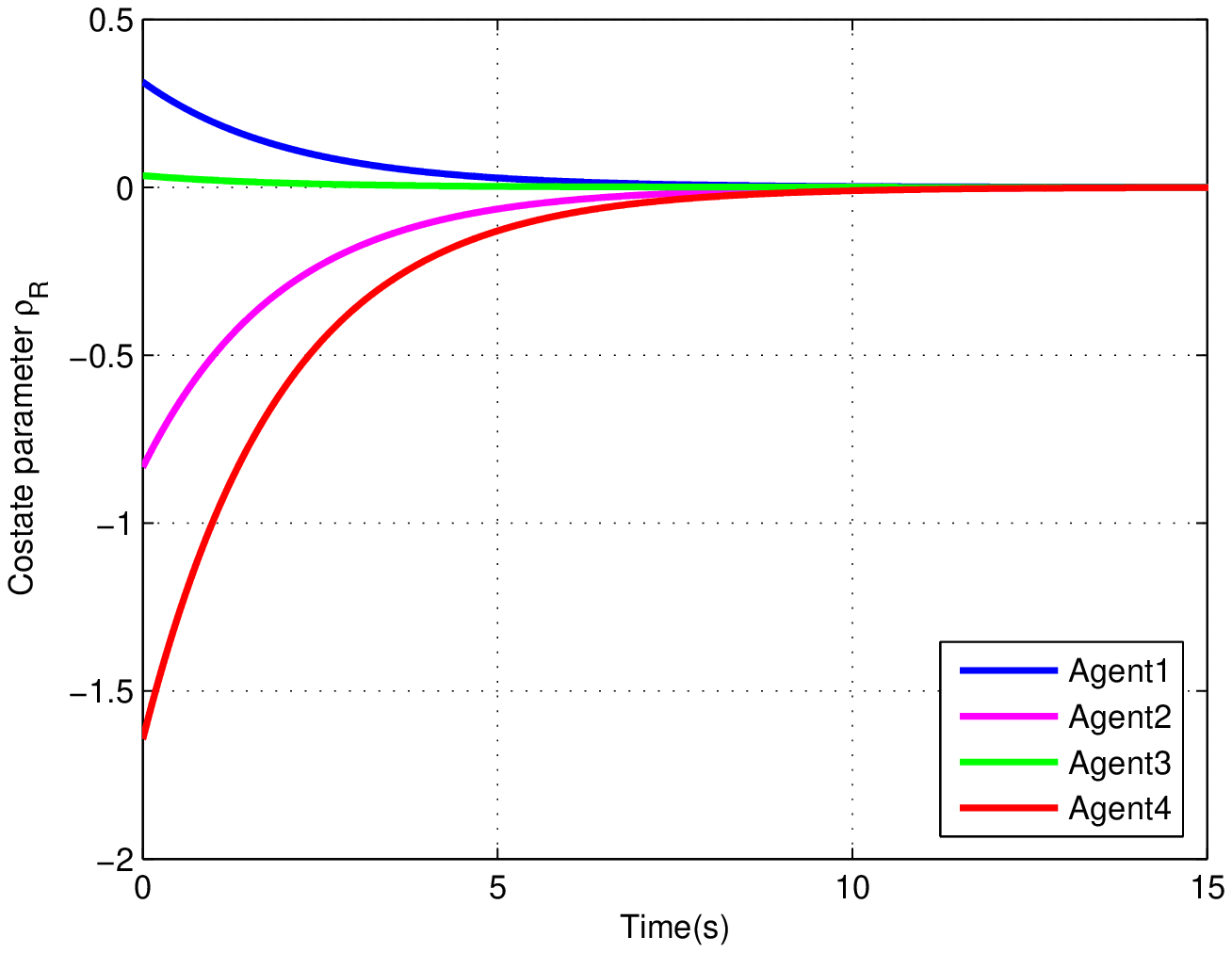}
\caption{Costate parameter $\rho_R$.\label{Fig11}}
  \label{Fig11}
\end{figure}

\begin{figure}[!hbt]
\centering
\includegraphics[width=3in]{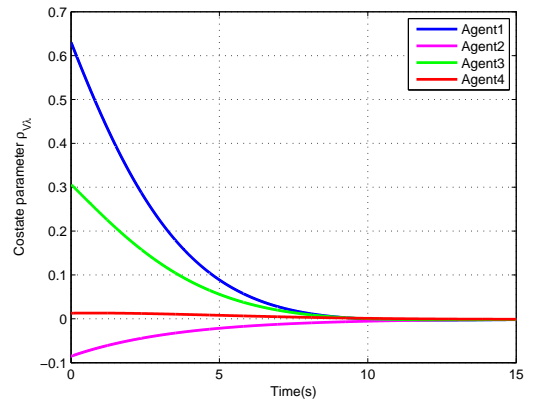}
\caption{Costate parameter $\rho_{V\lambda}$.\label{Fig12}}
  \label{Fig12}
\end{figure}


In this example, the acceleration component of the target is known, that is, the acceleration component along the direction of the target velocity is zero ($a_{tr}=0$ km/$s^2$), and the acceleration component perpendicular to the direction of the target velocity is time-varying ($a_{t\lambda}=0.1sin(10t)$ km/$s^2$). Then the acceleration components along and perpendicular to the LOS of the attackers are ${A_{Tr}}={a_{tr}}\cos{{\phi}}-a_{t\lambda}\sin{\phi}$ and ${A_{T\lambda}}=a_{tr}\sin{\phi}+a_{t\lambda}\cos{\phi}$. Attention should be paid to the fact that the accelerations of the target and the attacker are perpendicular to their respective velocity directions, which means that the speeds of the target and the attacker are constant and their direction are variable. The initial speeds of the target and the attacker are $V_i=0.7$ (km/s) and $V_T=1$ (km/s). In this example, four low-speed attackers attack a high-speed target at the same time. The guidance law in equation~(\ref{eq:3}) is adopted, in which matrices $P_{1i}=P_{2i}=I_N$, $t_0=0$(s), $t_f=15$(s), $R_0$ and $R_f$ are listed in Table 1. The initial values of costate parameters are $\rho_R(t_0)=V_r(t_0)$ and $\rho_{V\lambda}(t_0)=\dot{V_\lambda}(t_0)$ based on equation~(\ref{eq:7}).


Figures 3-12 depict the relative distance $R$, trajectory, relative velocity component along LOS $V_r$, relative velocity component perpendicular to LOS $V_\lambda$, input values ${A_{Tr}}$ and ${A_{T\lambda}}$, LOS angle $\lambda$ and costate parameters $\rho_R$ and $\rho_{V\lambda}$ of four low-speed attackers to strike a high-speed target at the same time. The simulation time is 15 seconds. It can be seen from the figures that the convergence of normal overload $\dot{V_\lambda}$ to zero makes the trajectory smooth, and the angular velocity of LOS $\dot{\lambda}$ is zero before the final strike, which means that multiple attackers can avoid internal collision in advance. The component of the attacker's acceleration along the LOS and the component of the attacker's acceleration perpendicular to the LOS have small chattering near zero. The reason is that the corresponding velocity component of acceleration control is zero, that is, when the velocity chatters near zero, the corresponding acceleration also has a small chattering near zero, and the acceleration components contain trigonometric functions ${A_{Mri}}=A_{Mi}\sin\xi_i$ and ${A_{M\lambda}}=-A_{Mi}\cos\xi_i$.

\subsection{Example 2: Cooperative attack with unknown target acceleration}

\begin{figure}[!hbt]
\centering
\includegraphics[width=3in]{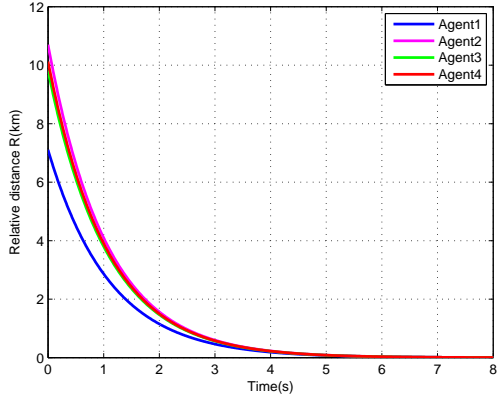}
\caption{Relative distances.\label{Fig13}}
  \label{Fig13}
\end{figure}

\begin{figure}[!hbt]
\centering
\includegraphics[width=3in]{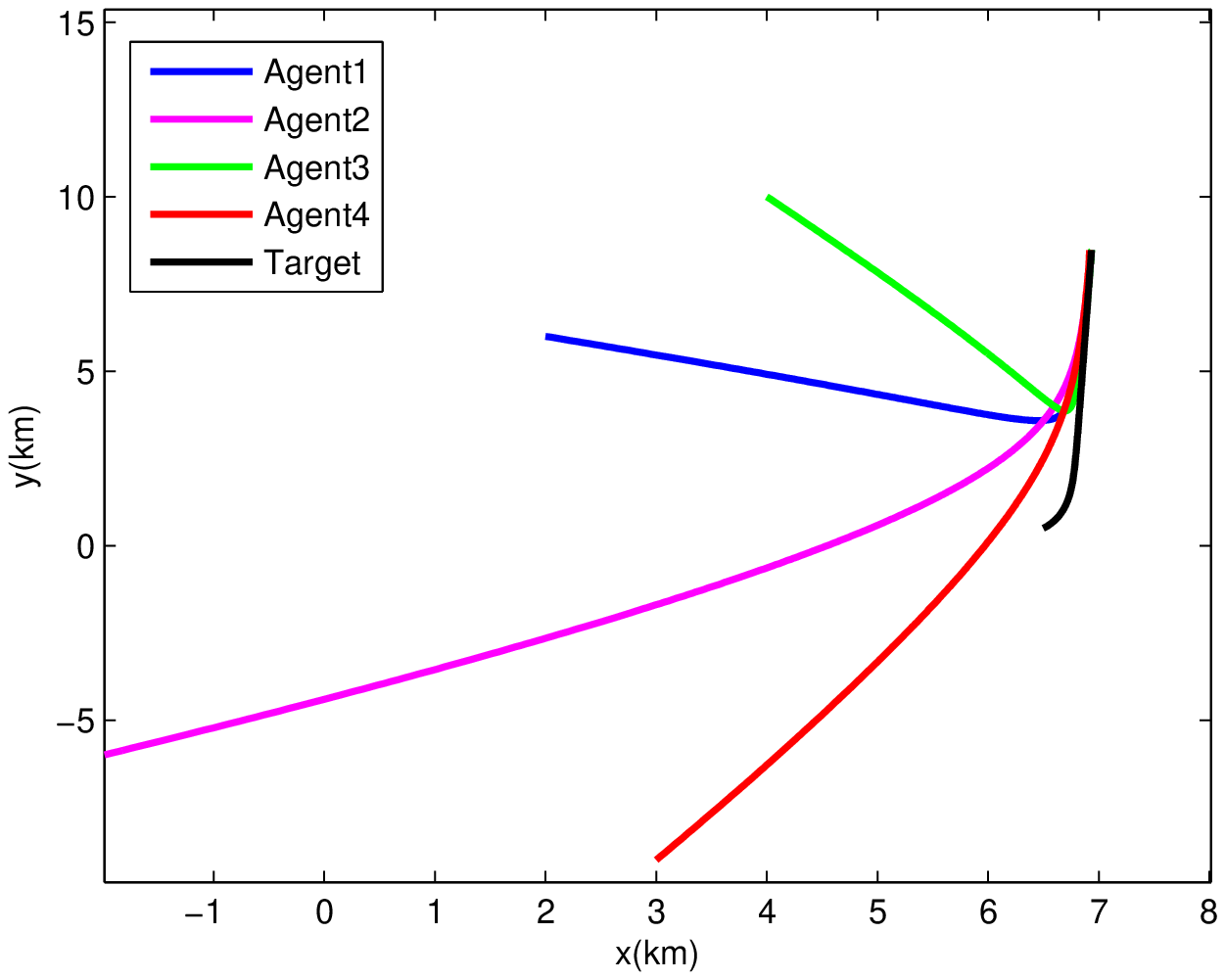}
\caption{Trajectories.\label{Fig14}}
  \label{Fig14}
\end{figure}

\begin{figure}[!hbt]
\centering
\includegraphics[width=3in]{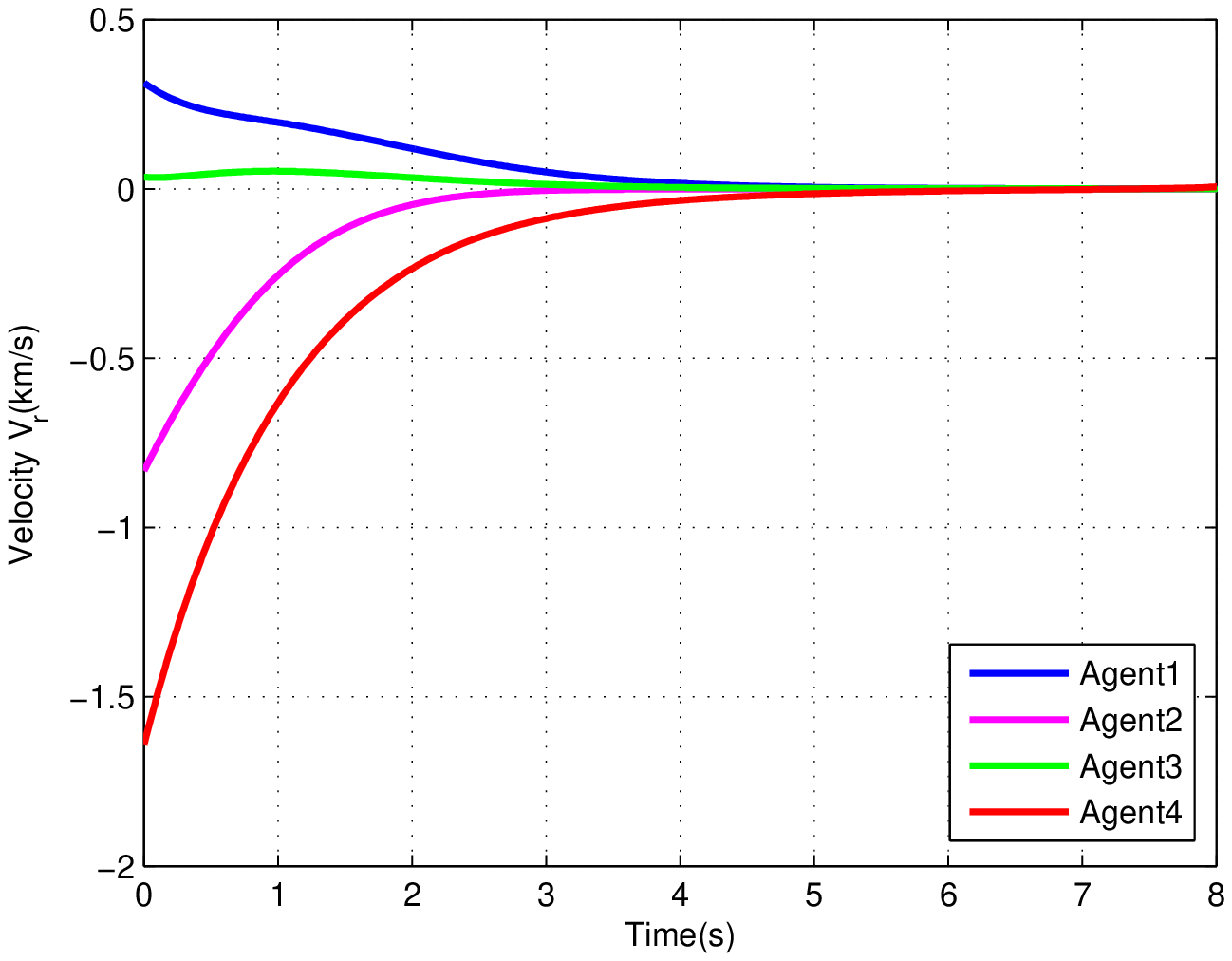}
\caption{Velocities ${V_r}$.\label{Fig15}}
  \label{Fig15}
\end{figure}

\begin{figure}[!hbt]
\centering
\includegraphics[width=3in]{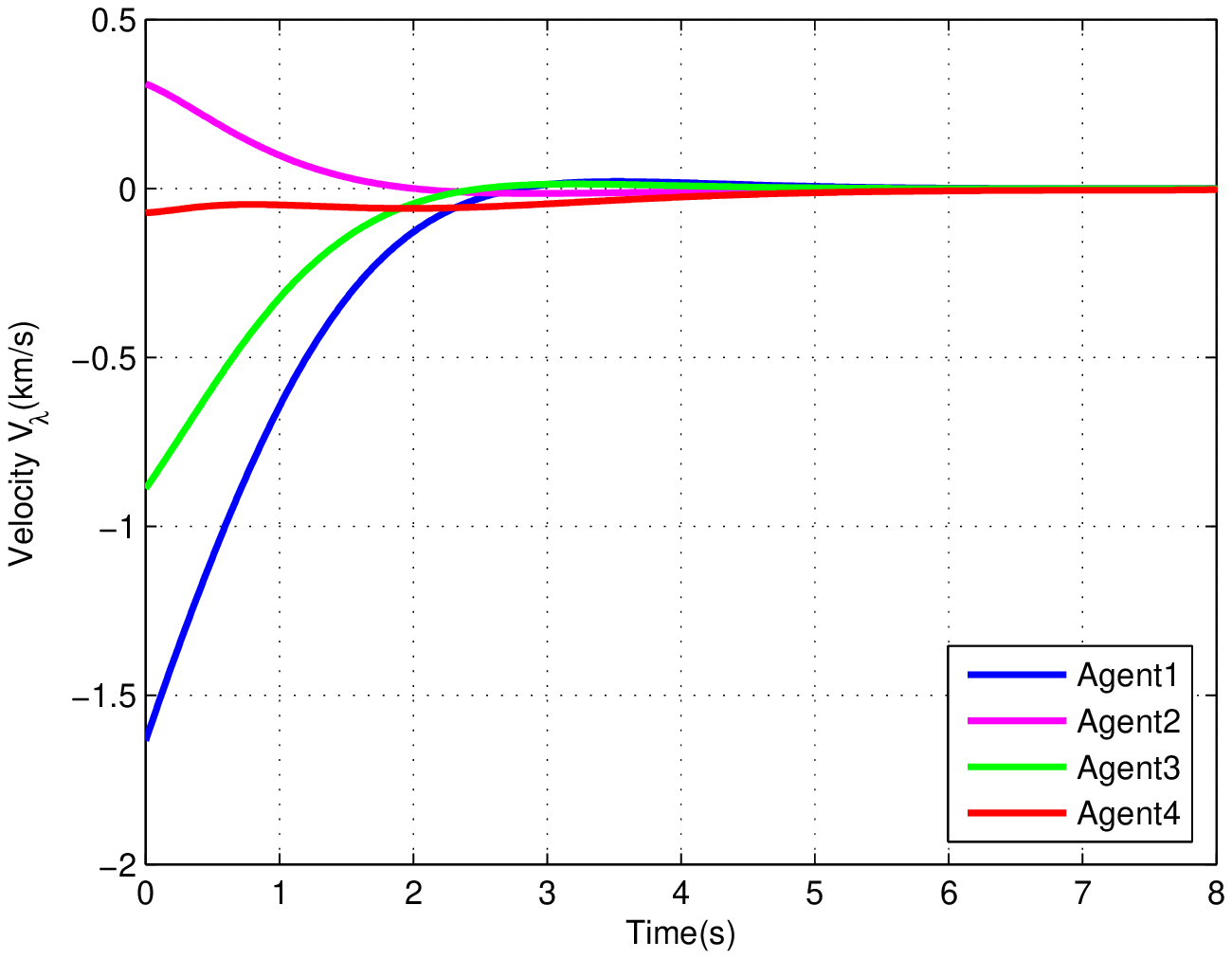}
\caption{Velocities ${V_\lambda}$.\label{Fig16}}
  \label{Fig16}
\end{figure}

\begin{figure}[!hbt]
\centering
\includegraphics[width=3in]{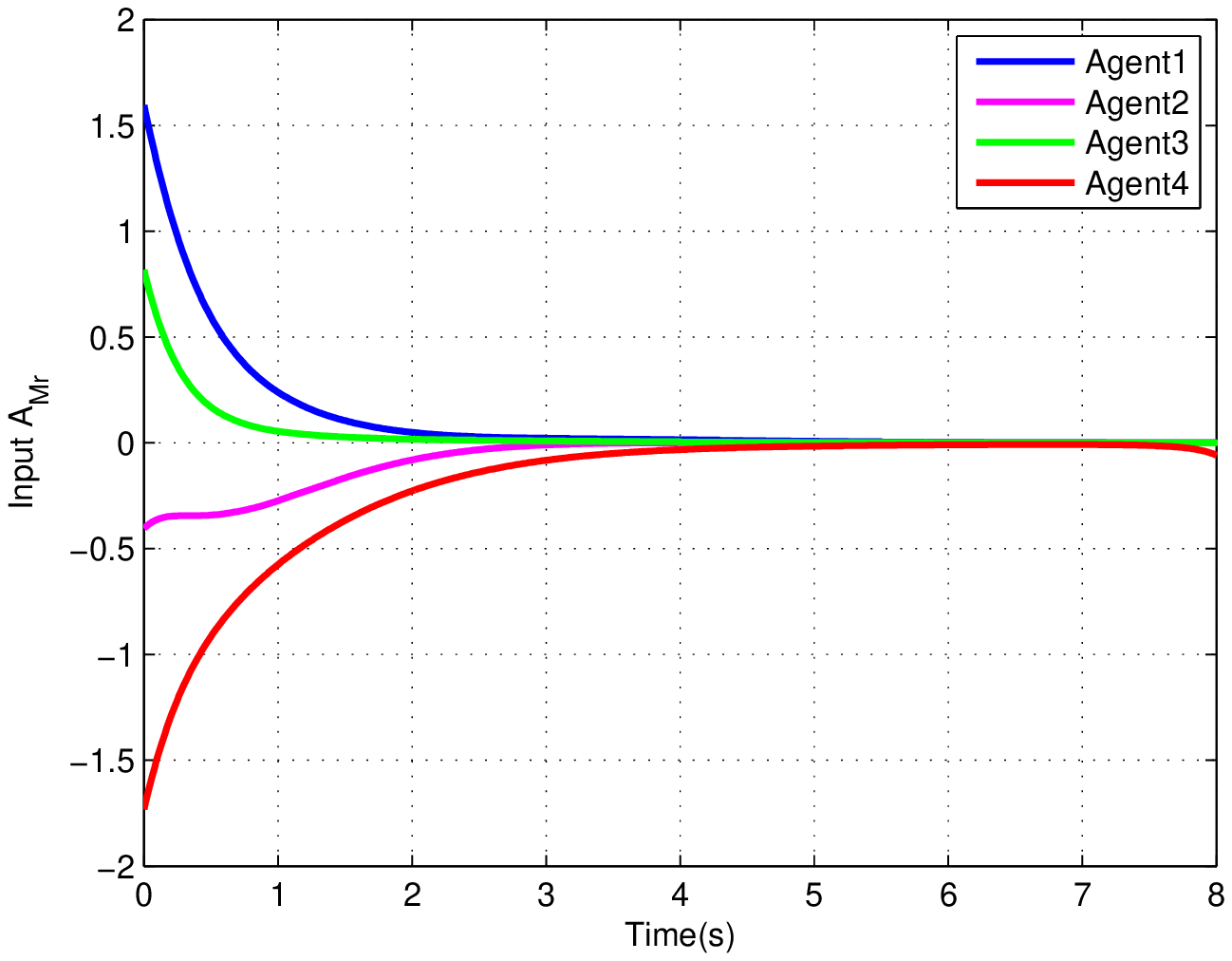}
\caption{Inputs ${A_{Mr}}$.\label{Fig17}}
  \label{Fig17}
\end{figure}

\begin{figure}[!hbt]
\centering
\includegraphics[width=3in]{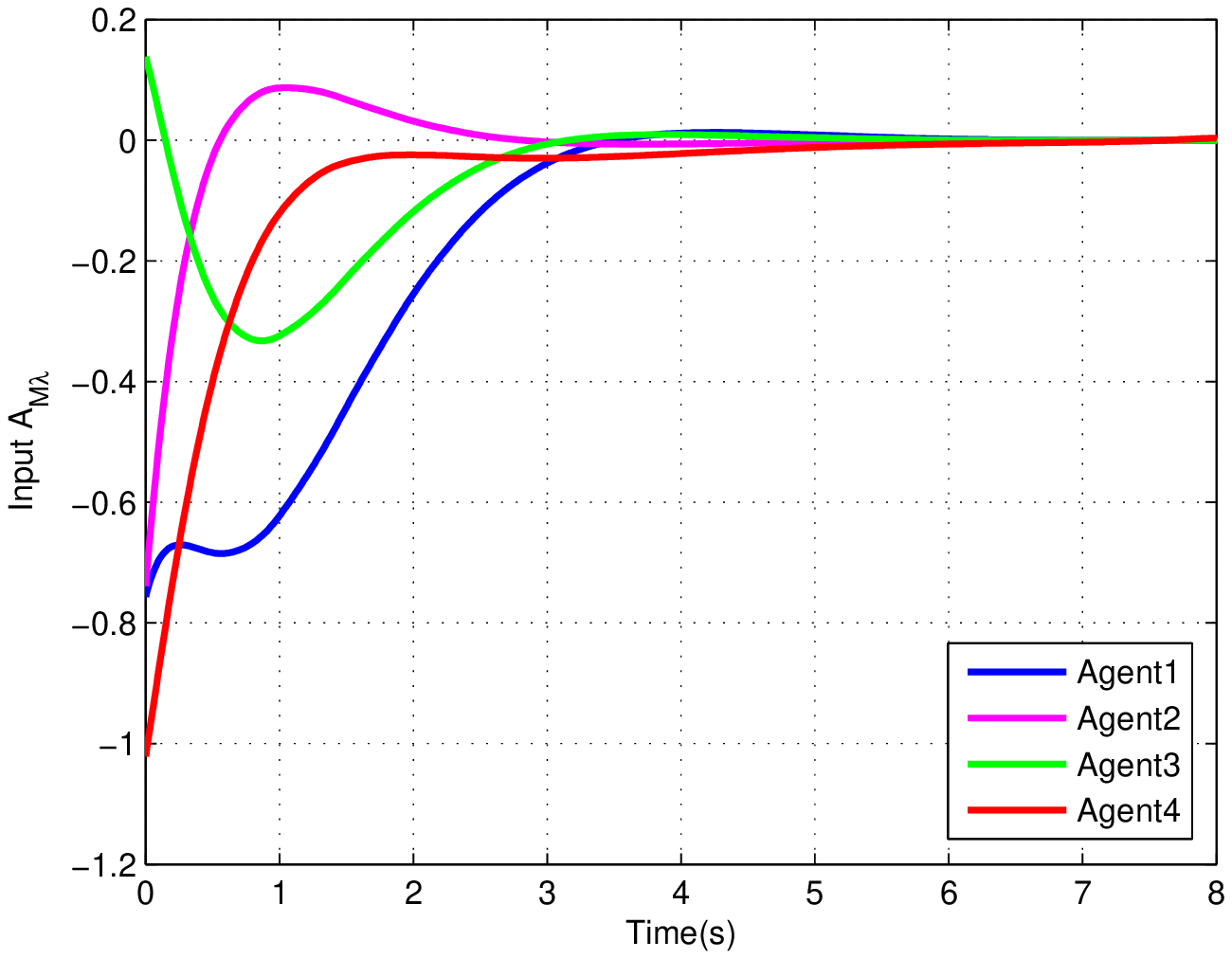}
\caption{Inputs ${A_{M\lambda}}$.\label{Fig18}}
  \label{Fig18}
\end{figure}

\begin{figure}[!hbt]
\centering
\includegraphics[width=3in]{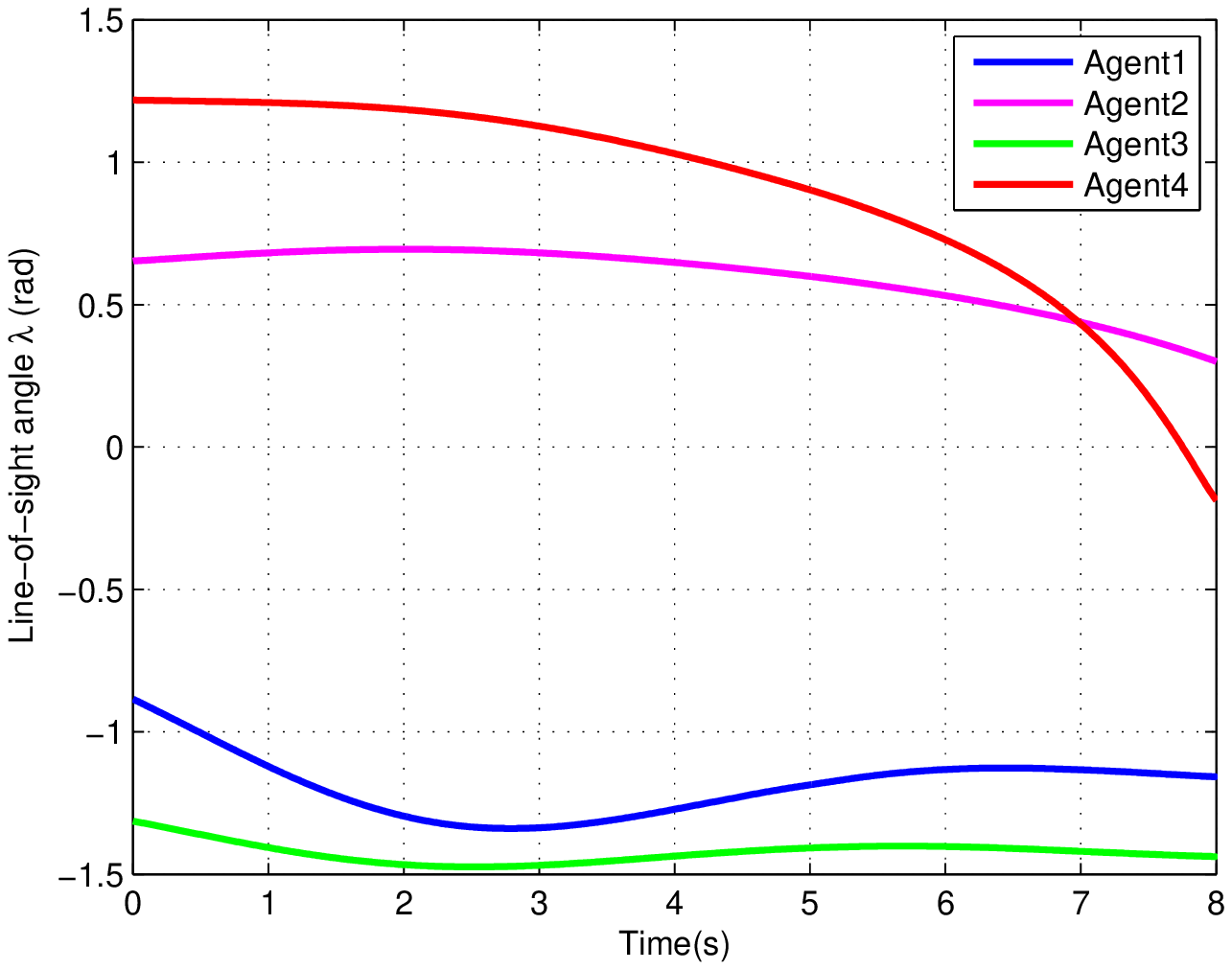}
\caption{Line of sight angle ${\lambda}$.\label{Fig19}}
  \label{Fig19}
\end{figure}

\begin{figure}[!hbt]
\centering
\includegraphics[width=3in]{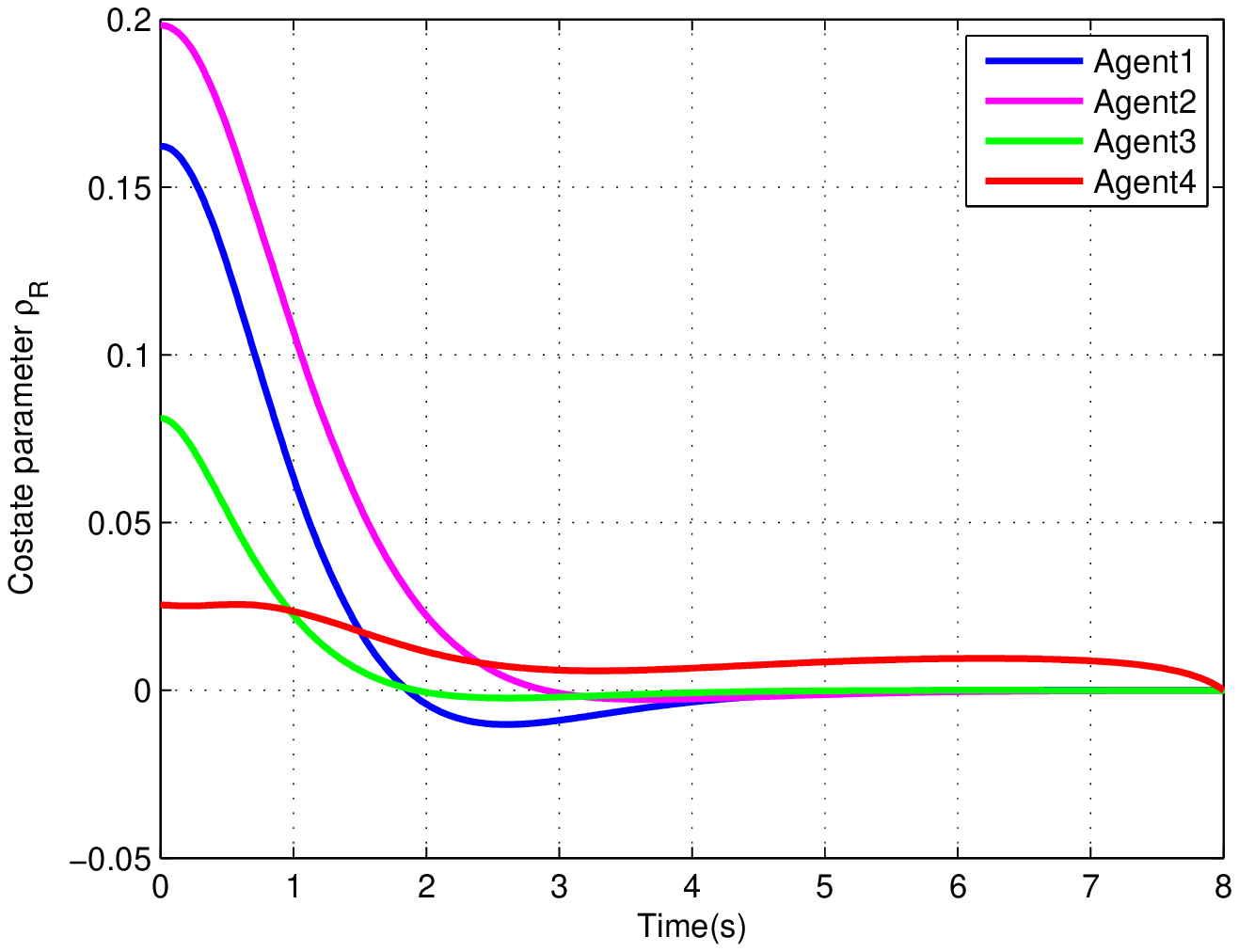}
\caption{Costate parameter $\rho_R$.\label{Fig20}}
  \label{Fig20}
\end{figure}

\begin{figure}[!hbt]
\centering
\includegraphics[width=3in]{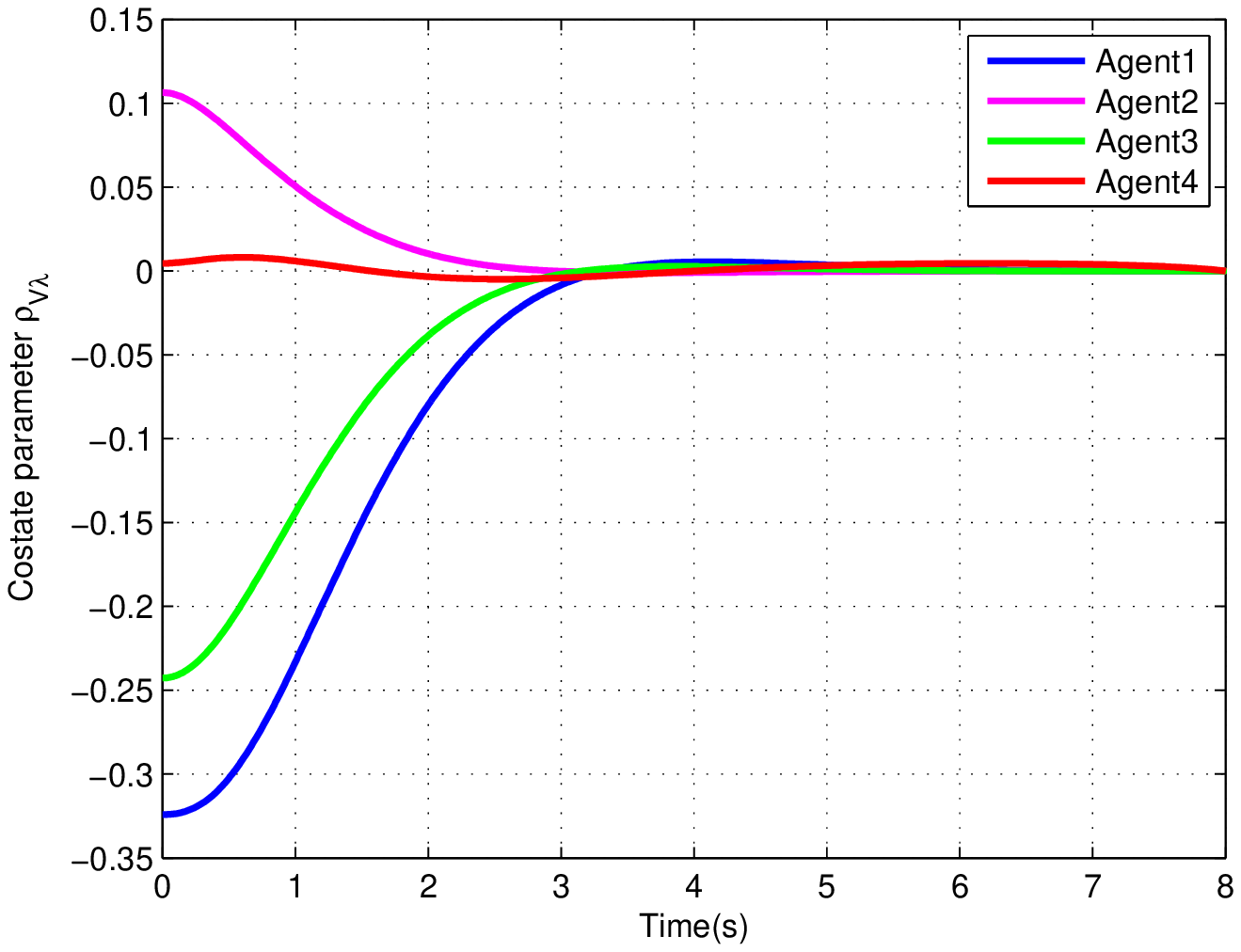}
\caption{Costate parameter $\rho_{V\lambda}$.\label{Fig21}}
  \label{Fig21}
\end{figure}

\begin{figure}[!hbt]
\centering
\includegraphics[width=3in]{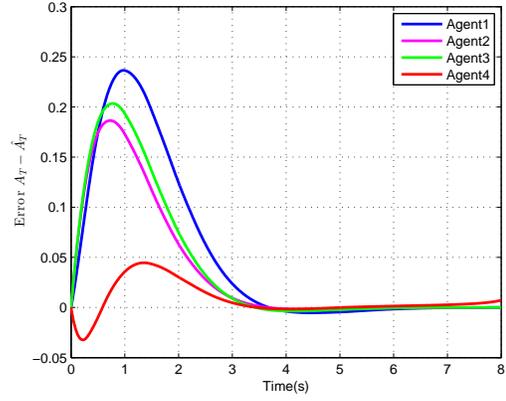}
\caption{Error $A_T-\hat{A_T}$.\label{Fig22}}
  \label{Fig22}
\end{figure}


In this case, the target acceleration is unknown and its structure is in equation~(\ref{eq:13}) where $s=-2$. The initial conditions of this example are the same as those listed above. The guidance law in equation~(\ref{eq:15}) is adopted, in which the parameters are the same as those listed above except $t_f=8$(s).


Figures 13-22 depict the relative distance $R$, trajectory, relative velocity component along LOS $V_r$, relative velocity component perpendicular to LOS $V_\lambda$, input values ${A_{Tr}}$ and ${A_{T\lambda}}$, LOS angle $\lambda$, costate parameters $\rho_R$ and $\rho_{V\lambda}$, and error $A_T-\hat{A_T}$ of four low-speed attackers to strike a high-speed target at the same time. The simulation time is 8 seconds. Similar to the above, the relative distance and relative speed of the attacker-target are consistent, and the simultaneous attack task can be completed in a limited time.

\section{Conclusions}

Distributed guidance laws based on two-point boundary value problem are designed in this paper, which can be used by multiple low-speed attackers to coordinate around or attack a high-speed moving target precisely at the same time. The acceleration of the target can be observed by an observer. At least one attacker can observe the information of the target, while the other attackers can obtain the information of the relative movement of the attacker-target indirectly from the communication network containing a directed spanning tree. The novel guidance laws can make the normal overload of the attacker-target relative motion zero, which means that the attacker's trajectory is smooth and meaningful. It can also fundamentally avoid the two difficulties in the design of multi-attacker simultaneous attacking target guidance law: calculation of the remaining time and avoidance of attackers' collision. In this paper, only the particle model of multi-UAV is studied, and its attitude control will be discussed in the future work.

\begin{table}
 \caption{\label{tab:table1}Simulation parameters of four attackers }
 \begin{tabularx}{9cm}{llllll}
  \hline
  $Parameters$ & ${Attacker}_1$ & ${ Attacker}_2$ & ${ Attacker}_3$ & ${ Attacker}_4$ \\  \hline 
   $\lambda_0,rad$    &-0.8851   & 0.6528  & -1.3135  &  1.2178 \\
   $\gamma_0,rad$    & 0.6283 &-1.0472 &-1.0472 &1.5708  \\
   $R_0,km$ & 7.1063 & 10.7005& 9.8234 &10.1242 \\
   $R_f,km$ & 0.0100 & 0.0100 & 0.0100 &0.0100 \\
   $V_{\lambda0},km/s$ & -1.6342  &  0.3099 &  -0.8881 &  -0.0722 \\
   $V_{\lambda f},km/s$ & 0.0100 & 0.0100 & 0.0100 &0.0100 \\
   \hline
  \end{tabularx} 
\end{table}

\begin{table}
 \centering
 \caption{\label{tab:table2}Simulation parameters of the target }
 \begin{tabularx}{6cm}{lll}
\hline
  $Target_x,km$ & $Target_y,km$ & $\sigma_T,rad$ \\ \hline 
   6.5000 & 0.5000 & 1.0472  \\
\hline
  \end{tabularx} 
\end{table}


%


\section*{Acknowledgment}
This research was supported by the National Natural Science Foundation of China under Grant No. 11332001 and No.61773024, Innovation Research Project Fund 17-163-11-ZT-003-018-01, and Joint Fund of the Ministry of Education of China 6141A020223.

\ifCLASSOPTIONcaptionsoff
  \newpage
\fi


\begin{thebibliography}{1}

\bibitem{1}
Zhou, J., and Yang, J., ``Distributed guidance law design for cooperative simultaneous attacks with multiple missiles,'' \textit{Journal of Guidance Control and Dynamics}, Vol.39, No.10, 2016, pp. 1--9.

\bibitem{2}
Hou, D., Wang, Q., Sun, X., and Dong, C., ``Finite-time cooperative guidance laws for multiple missiles with acceleration saturation constraints,'' \textit{Control Theory and Applications Iet}, Vol.9, No.10, 2015, pp. 1525--1535.

\bibitem{3}
Wang, X., Zhang, Y., and Wu, H., ``Distributed cooperative guidance of multiple anti-ship missiles with arbitrary impact angle constraint,'' \textit{Aerospace Science and Technology}, Vol.46, 2015, pp. 299--311.

\bibitem{4}
Zhou, J., Yang, J., and Li, Z., ``Simultaneous attack of a stationary target using multiple missiles, a consensus-based approach,'' \textit{Science China Information Sciences},  Vol.60, No.7, 2017, pp. 67--80.

\bibitem{5}%
Zhao, J., and Zhou,R., ``Unified approach to cooperative guidance laws against stationary and maneuvering targets,'' \textit{Nonlinear Dynamics}, Vol.81, No.4, 2015, pp. 1635--1647.


\bibitem{6}
Ren, W., and Beard, R.~W., ``Consensus seeking in multi agent systems under dynamically changing interaction topologies,'' \textit{IEEE Trans. on
  Automatic Control},  Vol.50, No.5, 2005, 655--661.

\bibitem{7}
Yu, W., Chen, G., Cao, M., and Kurths, J., ``Second-order consensus for multiagent systems with directed topologies and nonlinear dynamics,'' \textit{IEEE Trans. on Systems Man and Cybernetics Part B Cybernetics A Publication of the IEEE Systems Man and Cybernetics Society},  Vol.40, No.3, 2010, 881--891.

\bibitem{8}
Li Z, Chen MZQ and Ding Z., ``Distributed adaptive controllers for cooperative output regulation of heterogeneous agents over directed graphs,'' \textit{Automatica} 2016; 68: 179-183. 



\bibitem{9}
Bing, Z., Zaini, A. H. B., and Xie, L., ``Distributed guidance for interception by using multiple rotary-wing unmanned aerial vehicles,'' \textit{IEEE Transactions on Industrial Electronics}, Vol.64, No.7, 2017, pp. 5648--5656.

  \bibitem{10}
Song, Q., Liu, F., Wen, G., Cao, J., and Yang, X., ``Distributed position-based consensus of second-order multiagent systems with continuous/intermittent communication,'' \textit{IEEE Trans. on Cybernetics}, Vol.47, No.8, 2017, pp. 1860--1871.



\bibitem{11}
Cho, N., and Kim, Y., ``Modified pure proportional navigation guidance law for impact time control,'' \textit{Journal of Guidance Control and Dynamics}, Vol. 39, No. 4, 2016, pp. 1--21.


\bibitem{12}
Kim, H. G., Cho, D., and Kim, H. J., ``Sliding mode guidance law for impact time control without explicit time-to-go estimation,'' \textit{IEEE Transactions on Aerospace and Electronic Systems}, Vol.55, No.1, 2019, pp. 236--250.

\bibitem{13}
Lee, C. H., Kim, T. H., and Tahk, M. J., ``Interception angle control guidance using proportional navigation with error feedback,'' \textit{Journal of Guidance Control and Dynamics}, Vol. 36, No. 5, 2013, pp. 1556--1561.


\bibitem{14}
Lee, C. H., Kim, T. H., and Tahk, M. J., ``Effects of time-to-go errors on performance of optimal guidance laws,'' \textit{IEEE Transactions on Aerospace and Electronic Systems}, Vol.51, No.4, 2015, pp. 3270--3281.

\bibitem{15}
Garcia, E., Casbeer, D. W., and Pachter, M., ``Active target defense using first order missile models,'' \textit{Automatica}, Vol. 78, 2017, pp.139--143.



\bibitem{16}
Alkaher, D., Moshaiov, A., and Or,Y., ``Guidance laws based on optimal feedback linearization pseudocontrol with time-to-go estimation,'' \textit{Journal of Guidance Control and Dynamics}, Vol.37, No.4, 2014, pp. 1298--1305.


\bibitem{17}
Ghapani, S., Rahili, S., and Ren, W., ``Distributed average tracking of physical second-order agents with heterogeneous unknown nonlinear dynamics without constraint on input signals,'' \textit{IEEE Trans. on Cybernetics}, 2018, advance online publication.


\bibitem{18}
Rahili, S., and Ren, W., ``Distributed convex optimization for continuous-time dynamics with time-varying cost functions,'' \textit{IEEE Trans. on
  Automatic Control}, Vol.62, No.4, 2017, pp. 1590--1605.

  \bibitem{19}
Gao, W., Jiang, Z.~P., Lewis, F.~L., and Wang, Y., ``Leader-to-Formation Stability of Multi-Agent Systems: An Adaptive Optimal Control Approach,'' \textit{IEEE Trans. on
  Automatic Control}, Vol.63, No.10, 2018, pp. 3581--3587.



\bibitem{20}
Kang, S., Wang, J., Li, G., Shan, J., and Petersen, I.R., ``Optimal cooperative guidance law for salvo attack: an MPC-based consensus perspective,'' \textit{IEEE Transactions on Aerospace and Electronic Systems}, Vol.54, No.5, 2018, pp. 2397--2410.

\bibitem{21}
Zhou, J., and Yang, J., ``Guidance Law Design for Impact Time Attack Against Moving Targets,'' \textit{IEEE Transactions on Aerospace and Electronic Systems}, Vol.54, No.5, 2018, pp.2580--2589.

\end{thebibliography}
\end{document}